\documentclass[preprint,authoryear,review,12pt]{elsarticle}

\pdfoutput=1

\usepackage{xtab}
\usepackage{amsmath}
\usepackage{graphicx}
\usepackage{booktabs}
\usepackage{enumerate}
\usepackage{natbib}
\usepackage{dcolumn}
\usepackage{float}
\usepackage{siunitx}
\usepackage{multicol}
\usepackage{array}

\include {psfig}

\newcolumntype{d}[1]{D{.}{.}{#1} }

\journal{Icarus}

\begin{document}

\begin{frontmatter}

\title{Defining the Flora Family:\\ Orbital Properties, Reflectance Properties and Age}

\author[lpl]{Melissa J$.$ Dykhuis}
\ead{dykhuis@lpl.arizona.edu}
\author[calvin]{Lawrence Molnar}
\author[calvin]{Samuel J. Van Kooten}
\author[lpl]{Richard Greenberg}

\address[lpl]{Lunar and Planetary Laboratory, University of Arizona, 1629 E. University Blvd., Tucson, AZ 85719, USA}
\address[calvin]{Calvin College, 3201 Burton St SE, Grand Rapids, MI 49546, USA}

\begin {abstract}
  The Flora family resides in the densely populated inner main belt, bounded in semimajor axis by the $\nu_6$ secular resonance and the Jupiter 3:1 mean motion resonance. The presence of several large families that overlap dynamically with the Floras (e.g., the Vesta, Baptistina, and Nysa-Polana families), and the removal of a significant fraction of Floras via the nearby $\nu_6$ resonance complicates the Flora family's distinction in both proper orbital elements and reflectance properties. Here we use orbital information from the Asteroids Dynamic Site (AstDyS), color information from the Sloan Digital Sky Survey (SDSS), and albedo information from the Wide-field Infrared Survey Explorer (WISE) to obtain the median orbital and reflectance properties of the Floras by sampling the core of the family in multidimensional phase space. We find the median Flora SDSS colors to be $a^*$ = 0.126 $\pm$ 0.007 and $i-z = -0.037 \pm 0.007$; the median Flora albedo is $p_\text{V}$ = 0.291 $\pm$ 0.012. These properties allow us to define ranges for the Flora family in orbital and reflectance properties, as required for a detailed dynamical study. We use the young Karin family, for which we have an age determined via direct backward integration of members' orbits, to calibrate the Yarkovsky drift rates for the Flora family without having to estimate the Floras' material properties. The size-dependent dispersion of the Flora members in semimajor axis (the ``V'' plot) then yields an age for the family of $910^{+160}_{-120}$ My, with the uncertainty dominated by the uncertainty in the material properties of the family members (e.g., density and surface thermal properties). We discuss the effects on our age estimate of two independent processes that both introduce obliquity variations among the family members on short (My) timescales: 1) the capture of Flora members in spin-orbit resonance, and 2) YORP-driven obliquity variation through YORP cycles. Accounting for these effects does not significantly change this age determination.

\end {abstract}


\end{frontmatter}

\section {Introduction} 

\begin{figure}
\begin{center}
\includegraphics [width=5.0in]{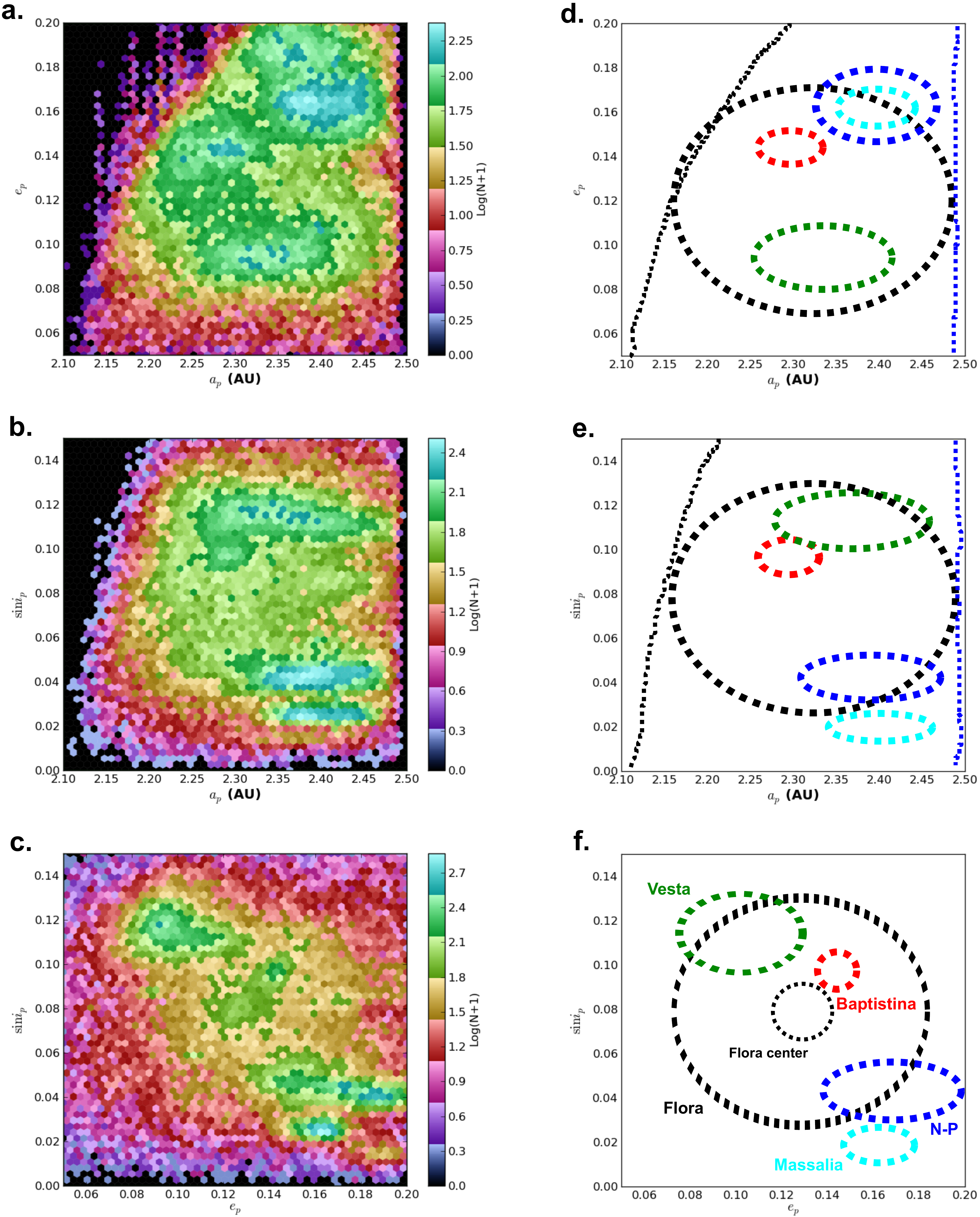}
\caption {\scriptsize Subplots a, b, and c show the number density of a zoomed-in subset of the 116,840 objects in Zone 1 (defined as the region for which 2.1 AU $< a < 2.5$ AU, $0 < e < 0.35$, $0 < \sin i < 0.3$) with an uncertainty parameter of two or less, according to the Minor Planet Center. We computed synthetic proper elements for these objects using the Orbit9 integration software from AstDyS. Subplots d, e, and f show the approximate outlines of the large groups typically identified in this orbital space (Flora = black, Vesta = green, Nysa-Polana = blue, Baptistina = red, Massalia = cyan), and the location of the $\nu_6$ secular resonance (black) and the Jupiter 3:1 mean motion resonance (blue).
\label{01_orbital}}
\end{center}
\end{figure}

The inner zone of the asteroid main belt is dominated by the large and diffuse Flora family (Figure \ref{01_orbital}). The first published mention of the Flora family \citep{hirayama1919} raised the difficulty of establishing the extent of the family. Since that early work, the possible single collision origin of the family \citep[][]{carusi1978,cellino1991,zappala1994,nesvorny2002} has been questioned in favor of a multi-collisional origin \citep[][]{brouwer1951,tedesco1979,zappala1990}, with the uncertainty in the origin models largely derived from the uncertainty in family membership. Membership determination is complicated by the dynamical overlap of several neighboring families, both large and small, with the Flora family (Figure \ref{01_orbital}).

\begin{figure}
\begin{center}
\includegraphics [width=3.7in]{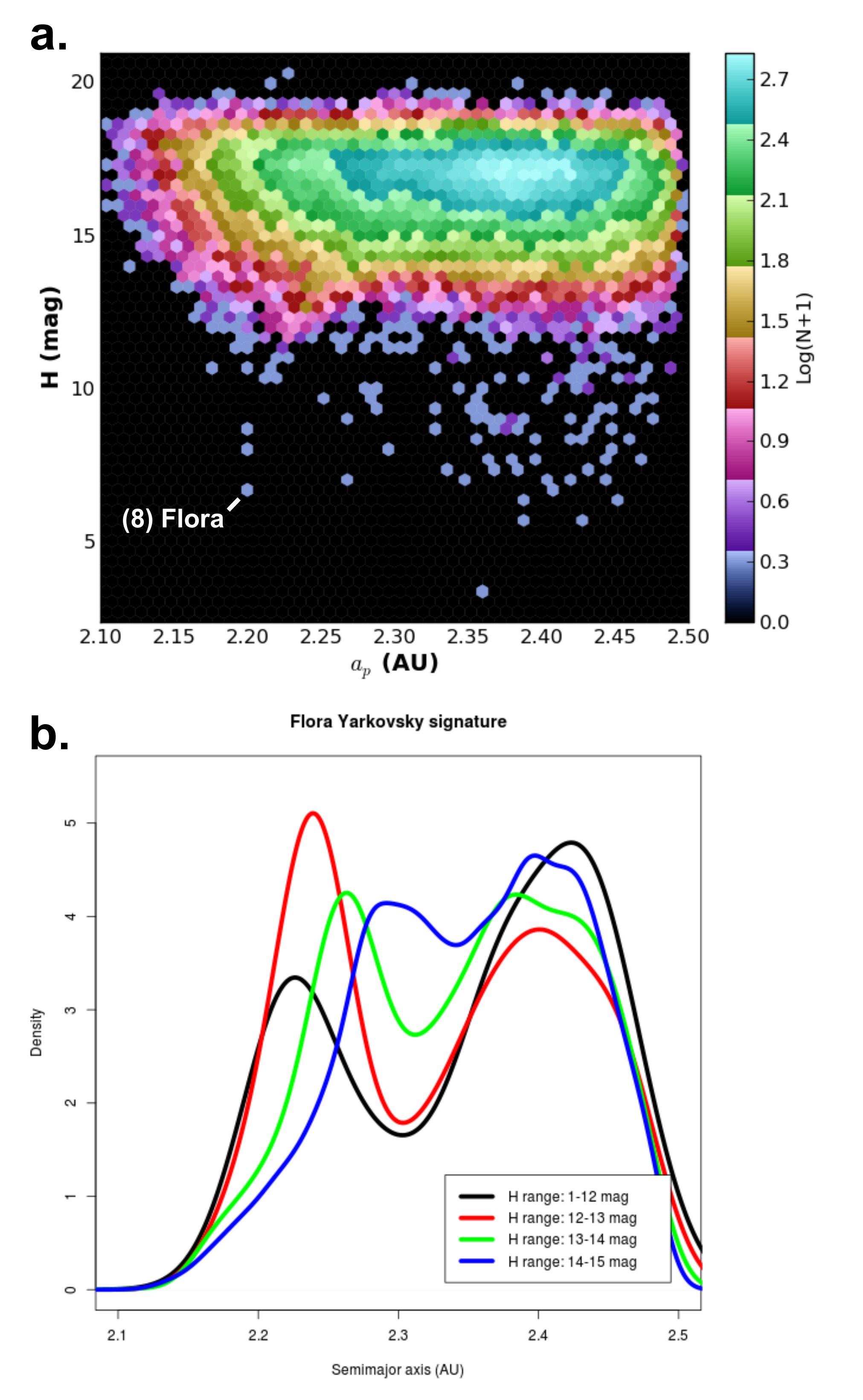}
\caption {\scriptsize \textbf{a)} Absolute magnitude and semimajor axis information for all 116,840 Zone 1 objects. The rightmost half of the Flora family's ``V'' signature is visible as a slight ridge branching toward higher $a$ and H from Flora's location at $a \sim$ 2.2 AU, H $\sim$ 6.5. \textbf{b)} Density curves (area under each curve is unity) for different absolute magnitude ranges in Zone 1. The Yarkovsky signature among the Flora members ($a$ from about 2.15 to 2.3 AU) manifests itself as a slow shifting of smaller objects toward higher semimajor axes. The largest objects (H from 1-12 mag) peak near 2.23 AU, objects with H from 12-13 mag peak near 2.25 AU, objects with H from 13-14 mag peak near 2.27 AU, and objects with H from 14-15 mag peak near 2.29 AU. The peaks visible near 2.4 AU are associated with the Nysa-Polana complex.
\label{02_aH}}
\end{center}
\end{figure}

The nearby strong $\nu_6$ secular resonance has dramatically shaped the Flora family's orbital distribution over its lifetime. Asteroid 8 Flora, the presumed largest remnant of the family's parent body, lies on the extreme inner edge of the family (at $a, e, \sin i$ of 2.20, 0.146, 0.098), with a current osculating perihelion distance of 1.858 AU, bringing it moderately close to Mars (aphelion distance 1.666 AU). If the location of 8 Flora is assumed to be near the semimajor axis location of the breakup of its parent body, it is plausible that the Flora family has lost most of its retrograde-spinning objects through drift via the Yarkovsky effect into the $\nu_6$ resonance \citep{vokrouhlicky2000,bottke2001}. The objects that remain, at higher semimajor axes, are primarily prograde rotators \citep[][discussed in Section \ref{spins} of this work]{haegert2009,hanus2013}, and form half of the characteristic ``V'' signature in plots of semimajor axis vs. size (Figure \ref{02_aH}), consistent with the Yarkovsky effect acting to disperse the prograde remnants of a single large-scale collision.

The proximity of the Flora family to the $\nu_6$ resonance also amplifies its role as a source region for the population of Mars-crossing and near-Earth objects (NEOs) as well as for Earth impactors in the past and future \citep{scholl1991,morbidelli2003,vernazza2008}. Determination of the family's characteristics, in particular, age and reflectance properties, will aid efforts to link the family with meteorite classes and constrain models for space weathering and the dynamical evolution of the asteroid belt.

These efforts are greatly assisted by recent additions to the existing observational data sets. New information includes the ongoing discovery and orbit characterization of small bodies via the Minor Planet Center Orbit Database (MPCORB\footnote{http://www.minorplanetcenter.net/iau/MPCORB.html}) and the Asteroids Dynamic Site (AstDyS\footnote{http://hamilton.dm.unipi.it/astdys/}); measurement of five-band colors via the Sloan Digital Sky Survey Moving Object Catalog \citep[SDSSMOC\footnote{http://www.astro.washington.edu/users/ivezic/sdssmoc/},][]{ivezic2002}; and determination of visual albedos via the Wide-field Infrared Survey Explorer \citep[WISE/NEOWISE\footnote{http://irsa.ipac.caltech.edu/Missions/wise.html},][]{wright2010,mainzer2011}. These data sets provide an unprecedented opportunity to constrain the collisional and dynamical history of the asteroid belt, using the multidimensional parameter space in which families of collisional origin tend to cluster \citep{parker2008,carruba2013,masiero2013}: not only dynamical parameters such as semimajor axis (\emph{a}), eccentricity (\emph{e}), and inclination (\emph{i}), but also reflectance properties such as absolute magnitude (H), visual albedo ($p_V$), and SDSS color parameters, especially the $a^*$ and $i-z$ colors defined below. 

This study focuses on the inner main belt objects for which $2.1 < a < 2.5$, $0 < e < 0.35$, and $0 < \sin i < 0.3$ (hereafter ``Zone 1''). Using the Orbit9 software available from AstDyS, we computed current synthetic proper elements for all of the objects within this zone that had reliable orbit determinations (uncertainty parameters of two or less, according to the Minor Planet Center's orbital uncertainty measurements). The proper elements are based on integrations of two million years for each object. Of those objects, we used only those that had reliable proper elements (approximately 10\% of the original catalog had less reliable proper elements due to resonant encounters with the planets). This yielded orbital proper element information for 116,840 objects within Zone 1.

The SDSSMOC provides reflectances of moving objects in five bands: $u$, $g$, $r$, $i$, and $z$, with effective wavelengths of 3540, 4770, 6222, 7632, and 9049 \AA{} and widths of 599, 1379, 1382, 1535 1370 \AA{} respectively \citep{fukugita1996}. The differences among families are most distinct in the linear combinations of $i-z$ and $a^*$, where $a^*$ was defined by \citet{ivezic2001} as

\begin{equation}
 a^* \equiv 0.89(g - r) + 0.45(r - i) - 0.57.
 \label{astar_eq}
\end{equation}

\noindent For the objects in the SDSS moving objects catalog that were observed more than once, we combined their observations in order to improve the uncertainties on individual measurements. We also revised uncertainty estimates based on a self-consistency check among the objects that had repeat observations, and rejected those with uncertainties greater than 0.1 magnitudes. The number of objects from our catalog of 116,840 in Zone 1 that have SDSS colors that meet this criterion is 25,351.

The WISE/NEOWISE data provide diameter measurements for over 100,000 asteroids \citep{masiero2011}, enabling the computation of their visual albedos via the relation \citep[e.g.,][]{fowler1992}

\begin{equation}
 D = \frac{1329}{\sqrt{p_\text{V}}}10^{-H/5},
 \label{pV_eq}
\end{equation}

\noindent where $H$ is the absolute magnitude. We obtained diameter measurements from WISE/NEOWISE, and combined these with the absolute magnitude data available from the Minor Planet Center to obtain the visual albedos of the objects observed by WISE/NEOWISE in Zone 1, rejecting those with albedo uncertainties greater than 0.1. The number of objects from our catalog of 116,840 in Zone 1 that have WISE albedos that meet this criterion is 15,504.

\begin{figure}
\begin{center}
\includegraphics [width=4.7in]{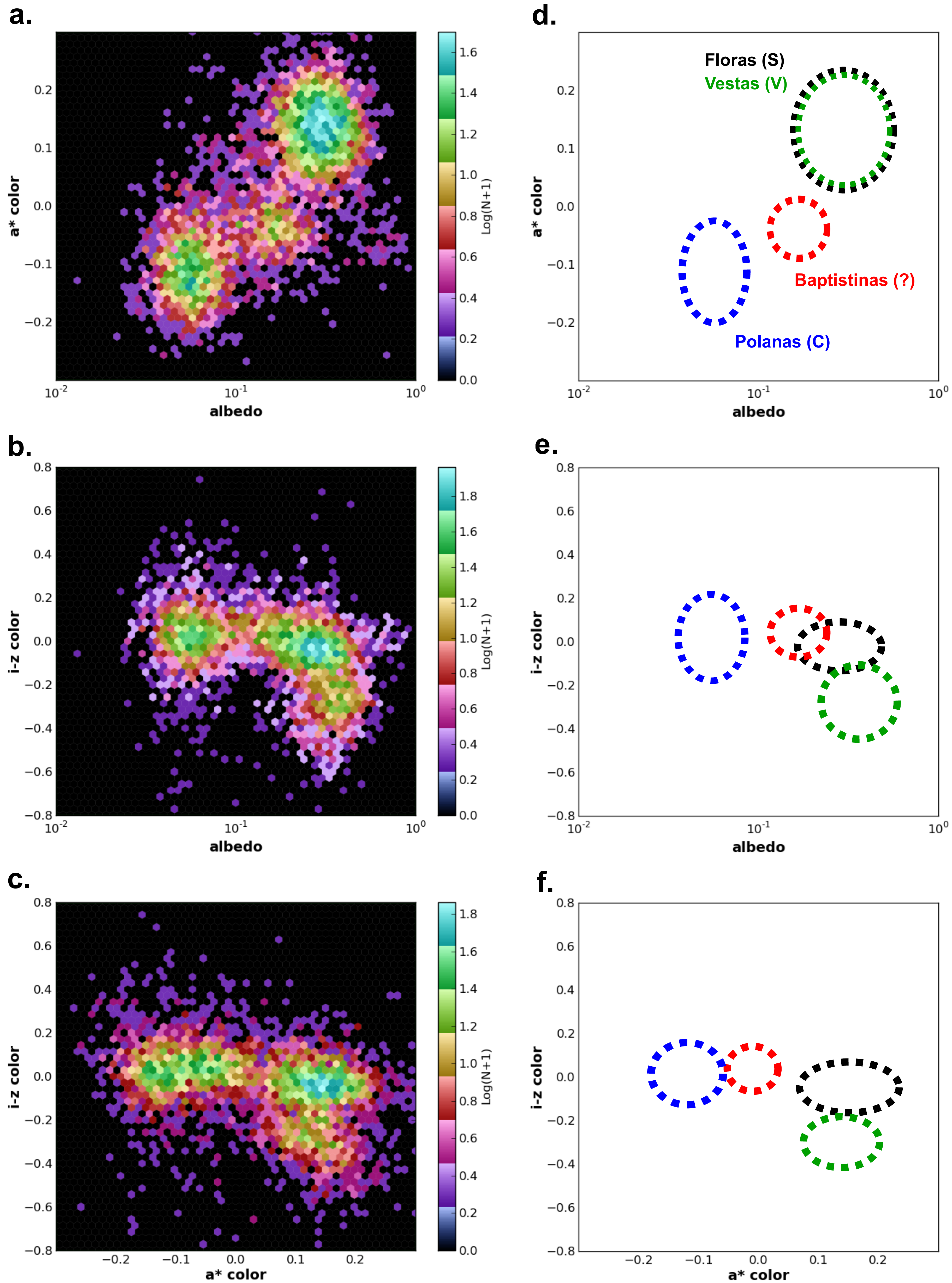}
\caption {\scriptsize \textbf{a, b, \& c:} Color distributions in three-dimensional space ($a^*$, $i-z$, $p_\text{V}$) for the 4696 Zone 1 objects observed by both SDSS and WISE. \textbf{d, e, \& f:} Families and spectral types associated with the clusters in (a,b,c): Floras (S-types) are outlined in black, Vestas (V-types) are outlined in green, Polanas (C-types) are outlined in blue, and Baptistinas (unclassified spectral type) are outlined in red. Two clusters located near $e$ and $\sin i$ of ($0.187, 0.040$) and ($0.170, 0.042$), historically associated with the Nysa clan and here determined to be related to objects 2751 Campbell and 135 Hertha respectively, overlap the Flora family in all three color-albedo dimensions.
\label{03_colors}}
\end{center}
\end{figure}

Both colors and albedos are known for a subset of 4696 Zone 1 asteroids observed by both SDSS and WISE. As shown in Figure \ref{03_colors}, they cluster into distinct groupings that correspond to standard spectral classifications. These groupings also correlate roughly with the families identified from the orbital element distribution in Figure \ref{01_orbital}, and these correlations are indicated by the labels in Figure \ref{03_colors}(d,e,f). In principle, the correlations can provide the means to identify members of a family despite dynamical overlap with other families and background objects. Such additional identification can allow improved study of the dynamical distribution of the family, yielding refined understanding of the nature of the initial family-forming event and the processes of subsequent orbital evolution. While the requirement of \emph{both} orbital and reflectance information significantly restricts the catalogs to the subset of those objects with both orbital and reflectance data, this restriction is not particularly injurious to the Flora family, due to its large numbers, and the combination of the datasets provides tremendous payoff in the disentangling of the overlapping families in Zone 1.

The association of the dynamical families in Zone 1 with particular spectral parameters has historically been rather blurry, due to the uncertainty of family membership for objects with overlapping dynamical parameters \citep{bendjoya1993,bendjoya2002}. This uncertainty has called into question the distinct spectral and dynamical properties of smaller families, such as the Baptistina family, which are dynamically buried within the Flora family. For example, in the recent past, Flora-like reflectance characteristics have been assigned to the Baptistina family \citep[e.g.,][]{reddy2011}, or the Baptistina family lumped together with the core of the Flora family \citep[referred to as the ``Belgica'' family by][]{carruba2013}. In addition to the Baptistina family, the Flora family overlaps with the large Vesta and Nysa-Polana populations, the three small families associated with 163 Erigone, 302 Clarissa, 752 Sulamitis, and additional smaller families (see Table \ref{neighborhood} in Section \ref{conclusion} for a complete list of overlapping families).

In this work, we set forward a procedure for family identification in which we minimize the effects of confusion by first focusing on the ``cores'' of families in both orbital and reflectance properties, where the fraction of interlopers from other families and background objects is assumed to be relatively small. Sampling the core in orbital proper elements enables us to identify the range of reflectance properties of a family; sampling the core of this range in reflectance properties enables us to identify the range of orbital elements of the family more precisely, and so on. This iterative procedure results in the definition of both the medians and ranges of orbital and reflectance properties of families, even in the cases of significant overlap. While the incompleteness of our catalogs precludes our obtaining complete family membership, which is necessary for extraction of collision information via the family's size-frequency distribution, the definition of the family's ranges in multidimensional space yields a sample of family members with a very low fraction of interlopers, ideal for further probes of the family's mineralogy and age.

In this paper, we apply the above techniques to the Flora family, demonstrating that the Flora family has well-defined ranges in both orbital proper elements and reflectance properties that distinguish it from nearby and overlapping families, and identifying the family as the product of a single collision dispersed by the Yarkovsky effect. In Section \ref{reflectance}, we use the iterative procedure described above to determine the median and range of the family's orbital and reflectance properties. This refined identification yields a more complete representation of the dynamical dispersion of the family, which allows us to correlate dispersion with asteroid size, placing constraints on orbital evolution due to the Yarkovsky effect over the time since the original family-forming impact event. From these constraints, we estimate the age of the Flora family via observations of the Yarkovsky spreading in semimajor axis (Section \ref{yarkovsky}). Finally, in Section \ref{discussion} we discuss the family's age and its remaining uncertainties, as well as implications for the spin distribution and mineralogy of the family members.

\section{Flora Median Orbital and Reflectance Properties}
\label{reflectance}

In this section, we define the median orbital and reflectance properties of the Flora family objects, and their ranges in color, albedo and proper elements. To obtain these definitions, we employ the following iterative procedure: we select objects from near the approximate center in orbital elements of the family and obtain from these the approximate reflectance properties of the family, then select objects from the center in reflectance space to obtain a refined estimate of the location of the dynamical center, etc. The end result of this analysis yields characteristic orbital and reflectance properties (i.e., the median values of $e$, $\sin i$, $a^*$, $i-z$, $p_\text{V}$, and $C(a)$, defined in Equation \ref{Cparam} below) for the Flora family, as well as ranges in orbital and reflectance parameters which describe the boundaries of the family. We also estimate the fraction of interlopers within the phase space we find to be occupied by the Floras. As the fraction is small, the phase space boundaries are objective and well constrained. The results of our analysis are given in Table \ref{flora_properties}, and an example of the steps of the iterative procedure are laid out in this section as follows:

\begin{enumerate}
 \item Locate the \emph{approximate} center in $e$ and $\sin i$ (Section \ref{approx_center}).
 \item Select a sample of objects dynamically near that center (Section \ref{objects_near_center}).
 \item Determine the reflectance properties of that sample; infer from these the \emph{approximate} reflectance properties of the family (Section \ref{approx_reflectance}).
 \item Select a sample of Zone 1 objects with reflectance properties close to those found in Section \ref{approx_reflectance}, with no restrictions in $e$ and $\sin i$ (Section \ref{floralike_reflectance}).
 \item Find the median orbital properties of the above sample, use these medians as a refined dynamical center (Section \ref{center_det}).
 \item Find the median reflectance properties of a sample of objects near the refined dynamical center (Section \ref{char_refl}).
 \item Define the ranges of the family in orbital and reflectance properties (Section \ref{membership}).
\end{enumerate}

The following seven subsections describe the steps of this process in detail.

To aid in the analysis required by this approach, we developed a powerful analysis tool, which we call ClusterAnalysis, for use in identifying family members from background objects on the basis of selections in multidimensional parameter space. More details of the ClusterAnalysis tool will be described in a follow-up paper (Molnar et al. in prep).

We note that the family spans the entire range of semimajor axes in Zone 1, and a concentrated core is not evident (nor physically meaningful, due to Yarkovsky dispersion) in that dimension alone. However, as is shown in Figure \ref{02_aH}, the range of the family in semimajor axis is defined by the ``V'' envelope that represents the limit of size-dependent semimajor axis drift due to the Yarkovsky effect on the Flora family members \citep{molnar2008,haegert2009}. Objects whose semimajor axes place them beyond this envelope can be assumed to be unrelated to the family. The boundary of this envelope is given by the limiting distance in semimajor axis $\Delta a$ from the assumed parent object 8 Flora \citep[from][]{vokrouhlicky2006}:

\begin{equation}
 \Delta a = C \cdot 10^{H/5}.
 \label{Cparam}
\end{equation}

\noindent The $C$ parameter thus enables us to define a ``center'' in semimajor axis around which the family clusters, and to define ranges in $C(a)$ that describe the outer bounds of the family. In the analysis that follows, we use the $C$ parameter as one of our dynamical parameters, as it allows us to set dynamically plausible limits on the family's semimajor axis range.

\subsection{Locate the approximate center in $e$ and $\sin i$}
\label{approx_center}

We begin our analysis with all of the 116,840 objects in Zone 1. By visual inspection (e.g., of Figure \ref{01_orbital}), the center of the Flora family in eccentricity and inclination lies near $e, \sin i$ of ($0.125, 0.080$). This first approximation is subjective, but it is refined in later steps.

\subsection{Select objects near the center}
\label{objects_near_center}

Next we determine the approximate reflectance properties of the family by sampling objects near this center, where contamination should be minimal. We select the 1009 objects with $0.120 < e < 0.130$, $0.075 < \sin i < 0.085$. The ranges here are chosen to avoid clear contamination from the Vesta, Baptistina and Nysa-Polana families.

In order to minimize interlopers further, we exclude from our sample any objects whose magnitude H and semimajor axis $a$ places it outside $-0.2$ mAU $< C < 0.2$ mAU. This removes 22 more objects, and the remaining 987 comprise our initial sample of objects that are ``Flora-like'' on the basis of their orbital properties.

\subsection{Determine the approximate reflectance properties}
\label{approx_reflectance}

\begin{figure}
\begin{center}
\includegraphics [width=3in]{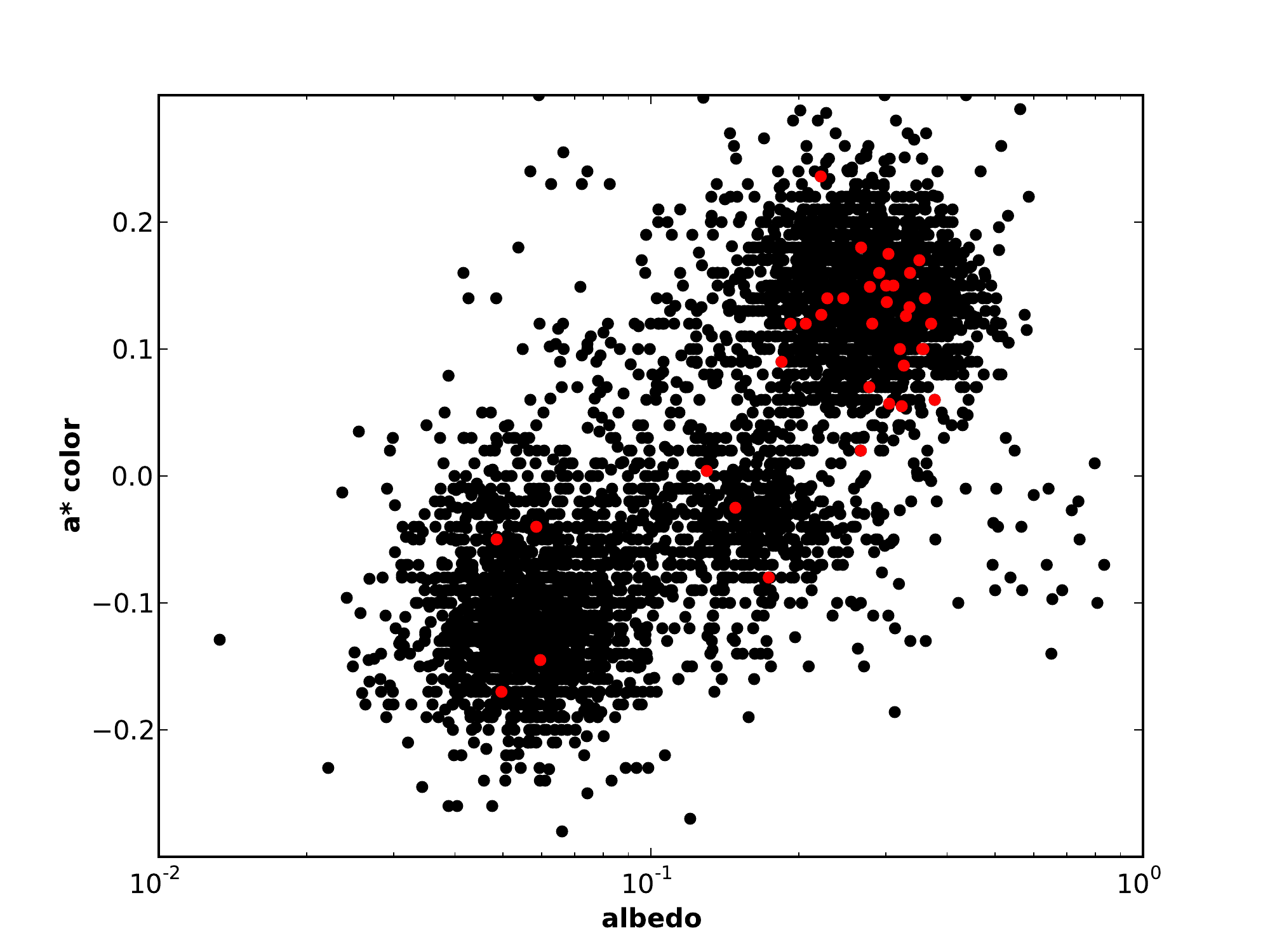}
\includegraphics [width=3in]{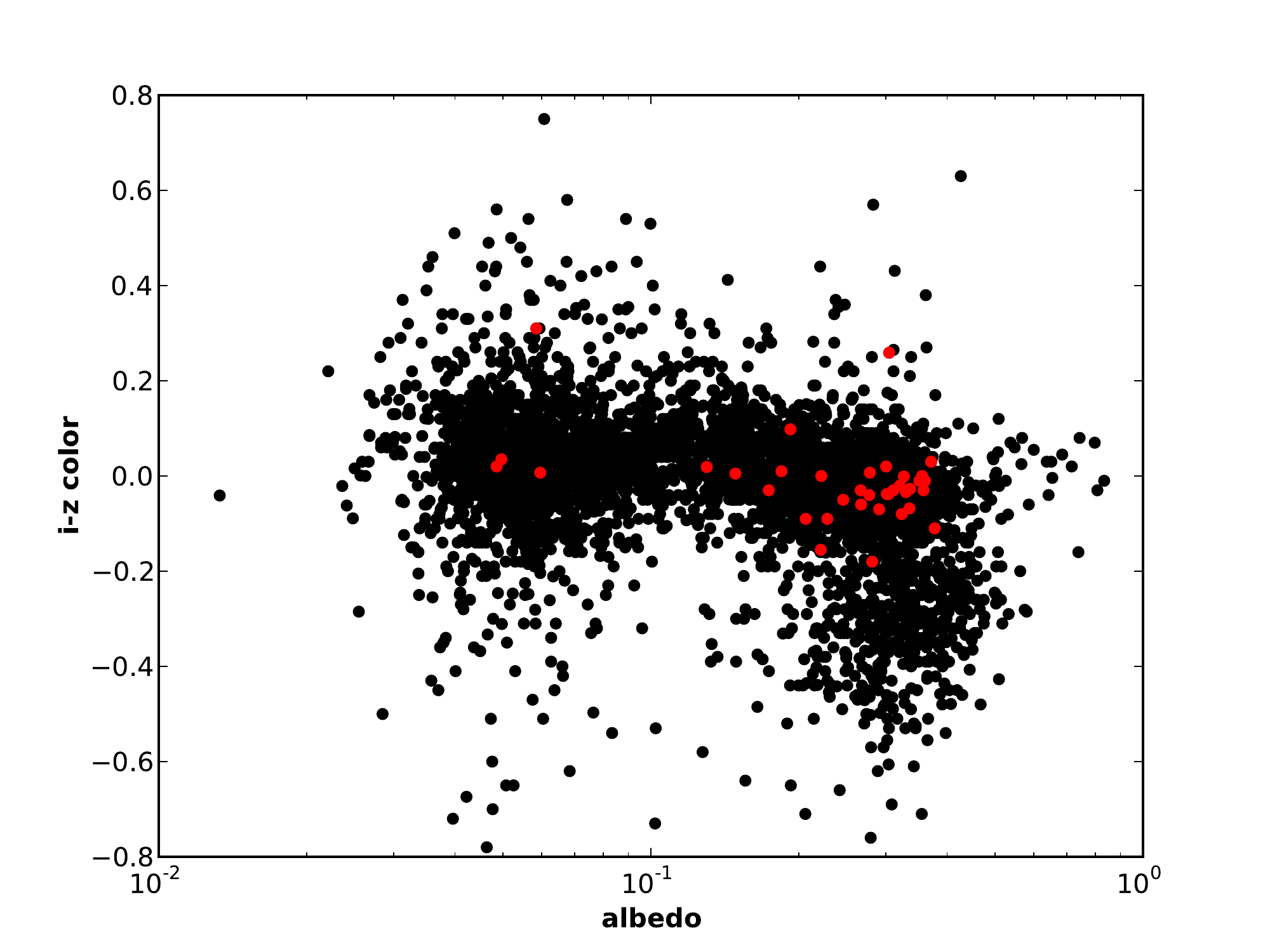}
\includegraphics [width=3in]{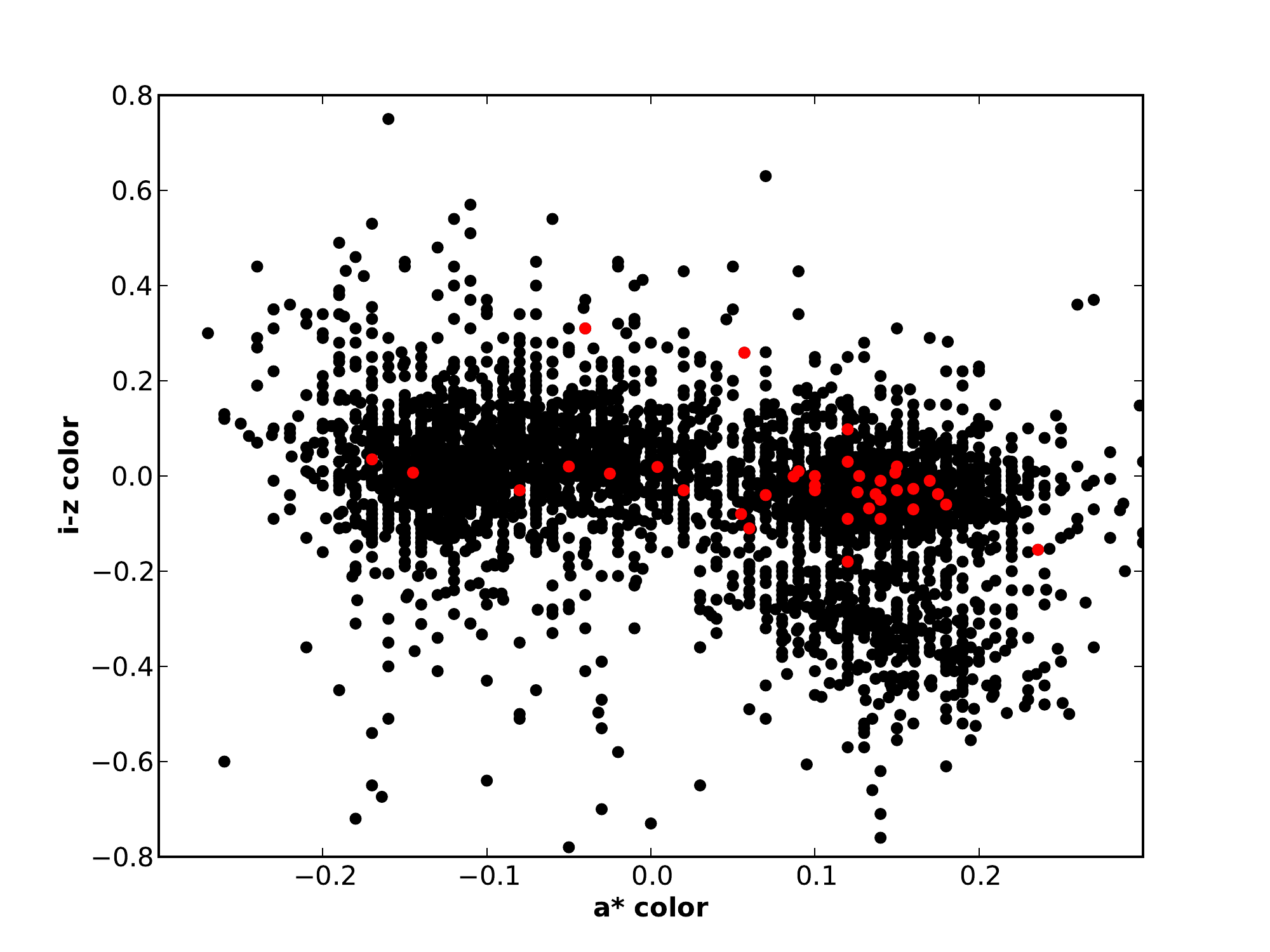}
\caption {Colors of the 37 objects observed by both SDSS and WISE among the 987 objects in the sample taken from the dynamical core (after selections in $e$, $\sin i$, and $C$). Background objects (black) are the 4696 Zone 1 objects observed by both SDSS and WISE.
\label{04_Floracolors}}
\end{center}
\end{figure}

Next we determine the median reflectance properties of the objects in this sample. Of the 987, we have both color (SDSS) and albedo (WISE) data for 37 of them, plotted in red in Figure \ref{04_Floracolors}. For comparison with the distribution in Figure \ref{03_colors}c, these points are plotted over a background (in black) of all 4696 objects from Zone 1 with color and albedo data. In Figure \ref{04_Floracolors}, we easily observe the two distinct groups (conventionally referred to as ``S'' and ``C'' types) among the background objects. The objects in our Flora sample (in red) contain a large majority that are consistent with S spectral type. 

The distributions plotted in red in Figure \ref{04_Floracolors} have median values of $a^*$, $i-z$, and $p_{\text{V}}$ of $0.12$, $-0.03$, and $0.28$, respectively. These values constitute our first approximation of the median reflectance properties of the Floras (to be refined in the next section).

\subsection{Select objects with Flora-like reflectance properties}
\label{floralike_reflectance}

\begin{figure}
\begin{center}
\includegraphics [width=6in]{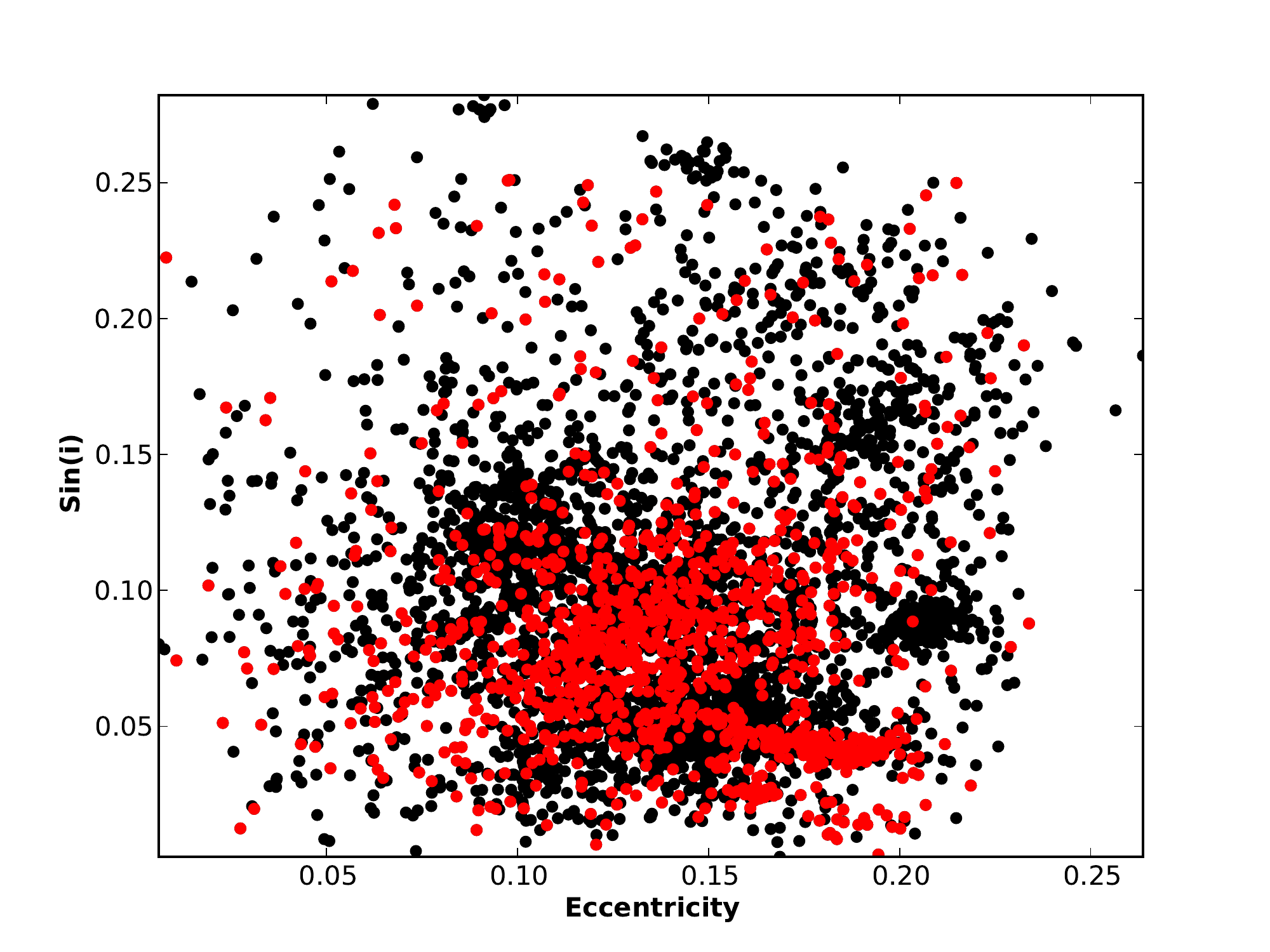}
\caption {Distribution in $e$ and $\sin i$ of objects with similar reflectance properties to the Flora family ($0.05 < a^* < 0.19$, $-0.15 < i-z < 0.09$, $0.20 < p_{\text{V}} < 0.40$). Due to the evident contamination from the Campbell and Hertha families near $e$ and $\sin i$ of ($0.187, 0.040$) and ($0.170, 0.042$) respectively, we must add additional restrictions in $a$ and $C$.
\label{05_Nysa_contamination}}
\end{center}
\end{figure}

Having identified median reflectance properties of objects near the dynamical center of the Flora family, we now select from Zone 1 all objects that have Flora-like reflectance properties (i.e., all of the objects near the median reflectance properties found in the previous section). Specifically, we select from the subset of 4696 objects observed by both SDSS and WISE those which have $0.05 < a^* < 0.19$, $-0.15 < i-z < 0.09$, and $0.20 < p_{\text{V}} < 0.40$. This section of parameter space is centered on the median reflectance properties of the Flora sample found in Section \ref{approx_reflectance}, with widths chosen to avoid contamination from the Vesta, Baptistina and Nysa-Polana families (note that the range in $p_{\text{V}}$ is centered in $\log_{10} p_{\text{V}}$). This subset contains 1152 objects. The $e$, $\sin i$ distribution of these objects is shown in red in Figure \ref{05_Nysa_contamination}. The distribution shows a background of diffuse objects throughout the space, upon which is centered a large cluster of Flora objects. In addition, the distribution shows clear contamination from additional families located near $e$ = 0.17 and $\sin i$ = 0.05 (discussed in the next section). 

\subsection{Determine the center in $e$ and sin$(i)$}
\label{center_det}

Next we use the distributions of objects with Flora-like reflectance properties to identify the Flora family's center in $e$ and $\sin i$ more precisely than the initial estimate obtained in Section \ref{approx_center}. A problem is presented by the irregular shape of the distribution in Figure \ref{05_Nysa_contamination}: the selections on the basis of color and albedo alone do not completely remove contamination from two nearby clusters of objects, centered at ($e$, $\sin i$) of ($0.187, 0.040$) and ($0.170, 0.042$). We find these clusters to be associated with asteroids 2751 Campbell and 135 Hertha, respectively; they have historically been assigned membership with the complicated ``Nysa-Polana'' complex. They are interlopers in the Flora clan, and skew the distribution of the family in $e$ and $\sin i$.

\begin{figure}
\begin{center}
\includegraphics [width=6in]{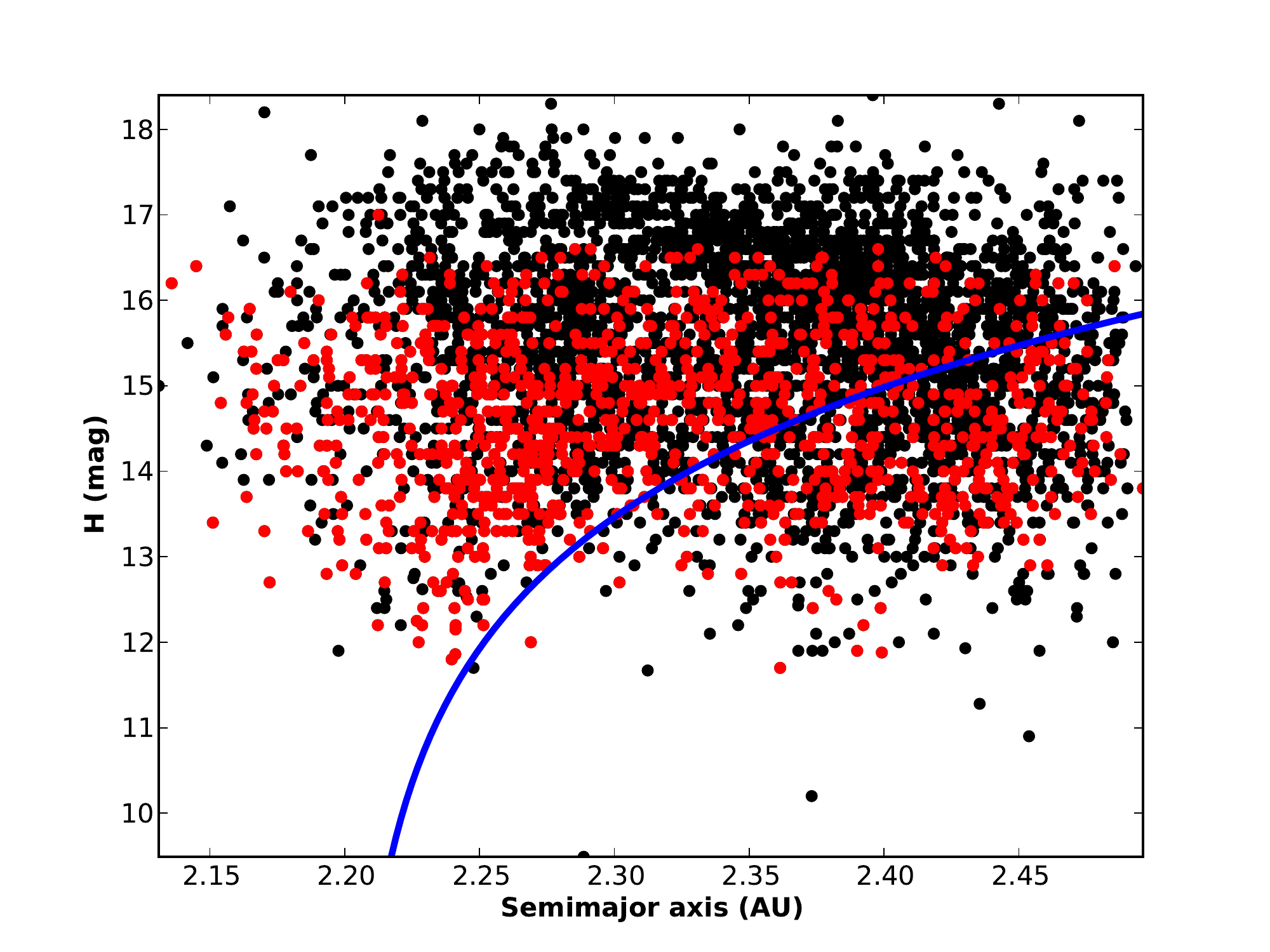}
\caption {Distribution in $a$ and H of objects with similar reflectance properties to the Flora family ($0.05 < a^* < 0.19$, $-0.15 < i-z < 0.09$, $0.20 < p_{\text{V}} < 0.40$). We restrict our sample to those objects within 2.1 AU $< a <$ 2.35 AU and $-0.2$ mAU $< C < 0.2$ mAU (blue line designates $C$ = 0.2 mAU).
\label{06_Ccut}}
\end{center}
\end{figure}

In order to eliminate these clusters, as in Section \ref{objects_near_center} we remove objects whose size and semimajor axis place them outside the family's Yarkovsky drift envelope. In addition, we eliminate objects with $a >$ 2.35 AU, in order to avoid significant contamination by Campbell and Hertha interlopers at higher semimajor axes. (This necessarily removes some Flora family members as well, but minimizes the interlopers in the sample, enabling us to obtain a distribution in $e$ and $\sin i$ that is more representative of the Flora family.) Figure \ref{06_Ccut} shows these limits in $a$, which reduce the number of objects under consideration to 678.

\begin{figure}
\begin{center}
\includegraphics [width=6in]{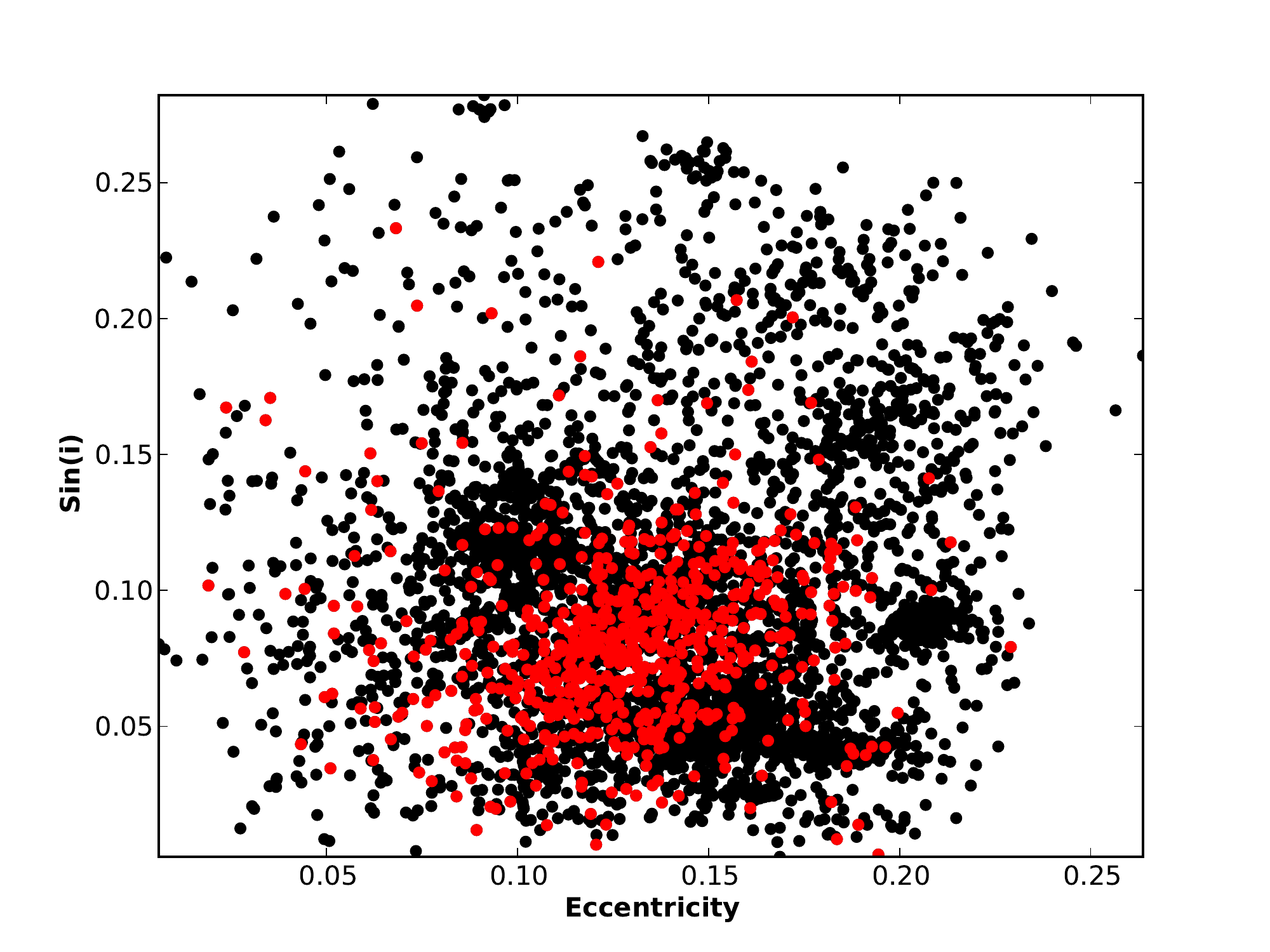}
\caption {As in Figure \ref{05_Nysa_contamination}, here we show the distribution in $e$ and $\sin i$ of objects with similar reflectance properties to the Flora family ($0.050 < a^* < 0.190$, $-0.150 < i-z < 0.090$, $0.195 < p_{\text{V}} < 0.395$). In addition, here we have restricted our sample to those objects within 2.1 AU $< a < 2.35$ AU and $-0.2$ mAU $< C < 0.2$ mAU. This extra restriction removes the contamination from the Campbell and Hertha families near $e$, $\sin i$ of ($0.187, 0.040$) and ($0.170, 0.042$) without biasing the distribution in $e$ and $\sin i$. The clear absence of these two families also demonstrates the importance of the selections made on $a$ and $C$ for removing interlopers from the family.
\label{07_noNysa}}
\end{center}
\end{figure}

The resulting sample in $e$ and $\sin i$ is shown in Figure \ref{07_noNysa}. Comparison with Figure \ref{05_Nysa_contamination} shows that the restrictions in semimajor axis largely eliminate the interlopers from the two clusters, reducing the population to a fairly symmetrical distribution. This symmetry allows us to refine the Flora family's center in $e$ and $\sin i$. The median $e$ of the objects in this sample is $0.130 \pm 0.002$, and the median $\sin i$ is $0.080 \pm 0.002$. (The uncertainties on the population medians were calculated according to the method described in Appendix \ref{appendixA}.) For comparison, recall from Section \ref{approx_center} that the original approximation of the center in $e$ and $\sin i$ was ($0.125, 0.080$). The slight correction is enough to have justified the effort, but is small enough to show that further iterations would be unlikely to yield significant improvement.

\subsection{Determine Flora characteristic reflectance properties}
\label{char_refl}

We next determine the median reflectance properties of the family by sampling the region near the refined center in $e$ and $\sin i$. In order to assess the sensitivity of the measured reflectance properties to the exact choice of center in $e$ and $\sin i$, we systematically explored the properties across a grid of possible ``centers'' within the range of uncertainty in $e$ and $\sin i$: $0.128 < e < 0.132$ and $0.078 < \sin i < 0.082$ (see Appendix \ref{appendixB} for further details of this procedure). The values reported in Table \ref{flora_properties} represent the median of all of the medians measured at each center within the grid, with the range of all possible values given as the uncertainty. 


As with the median orbital elements derived in Section \ref{center_det}, the results for the reflectance properties reported here show slight differences from the original approximation in Section \ref{approx_reflectance}, enough to justify the effort, yet small enough that further iterations are not needed. 

We take the median values reported in Table \ref{flora_properties} to be the characteristic orbital and reflectance properties of the Flora family, which allow us to distinguish it from overlapping families. In the next section we define ranges in orbital elements and reflectance properties within which the Flora family resides, to allow us to probe the mineralogy and age of the family in the later sections.

\begin{table}[H]
{\small
\begin{tabular}{ | l | >{\bfseries}r c | c | c | }
 \hline 
 Parameter        & median   & median unc. range   & literature        & sample range \\ [1.0ex]
\hline
 $a \text{(AU)}$  &          &                     &                   & 2.1 to 2.5      \\
 $e$              & 0.130    & 0.128 to 0.132      & 0.133             & 0.065 to 0.19   \\
 $\sin i$         & 0.080    & 0.078 to 0.082      & 0.084             & 0.025 to 0.13   \\
 $C \text{(mAU)}$ &          &                     &                   & -0.2 to 0.2     \\
 $a^*$            & 0.126    & 0.120 to 0.133      & 0.13              & 0.055 to 0.22   \\
 $i-z$            & -0.037   & -0.040 to -0.030    & -0.05             & -0.14 to 0.090  \\
 $p_{\text{V}}$   & 0.291    & 0.279 to 0.299      & 0.288 $\pm$ 0.088 & 0.17 to 0.40    \\
\hline
\end{tabular}
}
 \caption{Flora family median orbital and reflectance properties, this work compared with published literature values. Due to the signatures of the Yarkovsky spreading and removal via the $\nu_6$ resonance, a median value for $a$ is not physically meaningful, and is not reported; however, both $a$ and $C$ provide useful metrics for selection of family members from the background in Section \ref{membership}. Published values for the family's characteristic $e$, $\sin i$, $a^*$ and $i-z$ are from \citet{parker2008}; the value for the family's albedo is from \citet{masiero2013}. In addition, \citet{broz2013} measured an albedo of 0.304 for the Floras. In the last column are the ranges that describe the distribution of the Flora family in orbital and reflectance properties (Section \ref{membership}).}
 \label{flora_properties}
\end{table}

\subsection{Define ranges of Flora sample}
\label{membership}

The hierarchical clustering method \citep[HCM, e.g.,][]{nesvorny2012} for family identification is based on a threshold distance in orbital element space of neighboring asteroids. However, a high fraction of interlopers due to overlapping families can connect unrelated regions of phase space for standard threshold values (e.g., assigning all asteroids within Zone 1 Flora family membership), while use of more conservative thresholds may fail to show the full extent of the physical family (e.g., identifying only the core of the Flora family as a dynamical family). Here we use additional dimensions to reduce the incidence of interlopers and focus instead on defining the ranges of the Flora family in proper orbital elements and reflectance properties. These well-defined ranges allow us to study the Yarkovsky dispersion, spin distribution, and spectra/mineralogy of the family more carefully, our primary goal. The incompleteness of the color and albedo catalogs upon which we rely in this work necessarily prohibits us from studying the unbiased size frequency distribution of the family, at least in the present work.

In order to determine the range of the family in a particular orbital or reflectance parameter, we look at the distribution of family members in that parameter. For each parameter, in order to ensure that the distribution is that of the family (and not the background or neighboring families), we select only the objects that are near the medians in all of the other parameters (and thus are expected to be mostly family members). 

\begin{figure}
\begin{center}
\includegraphics [width=5.2in]{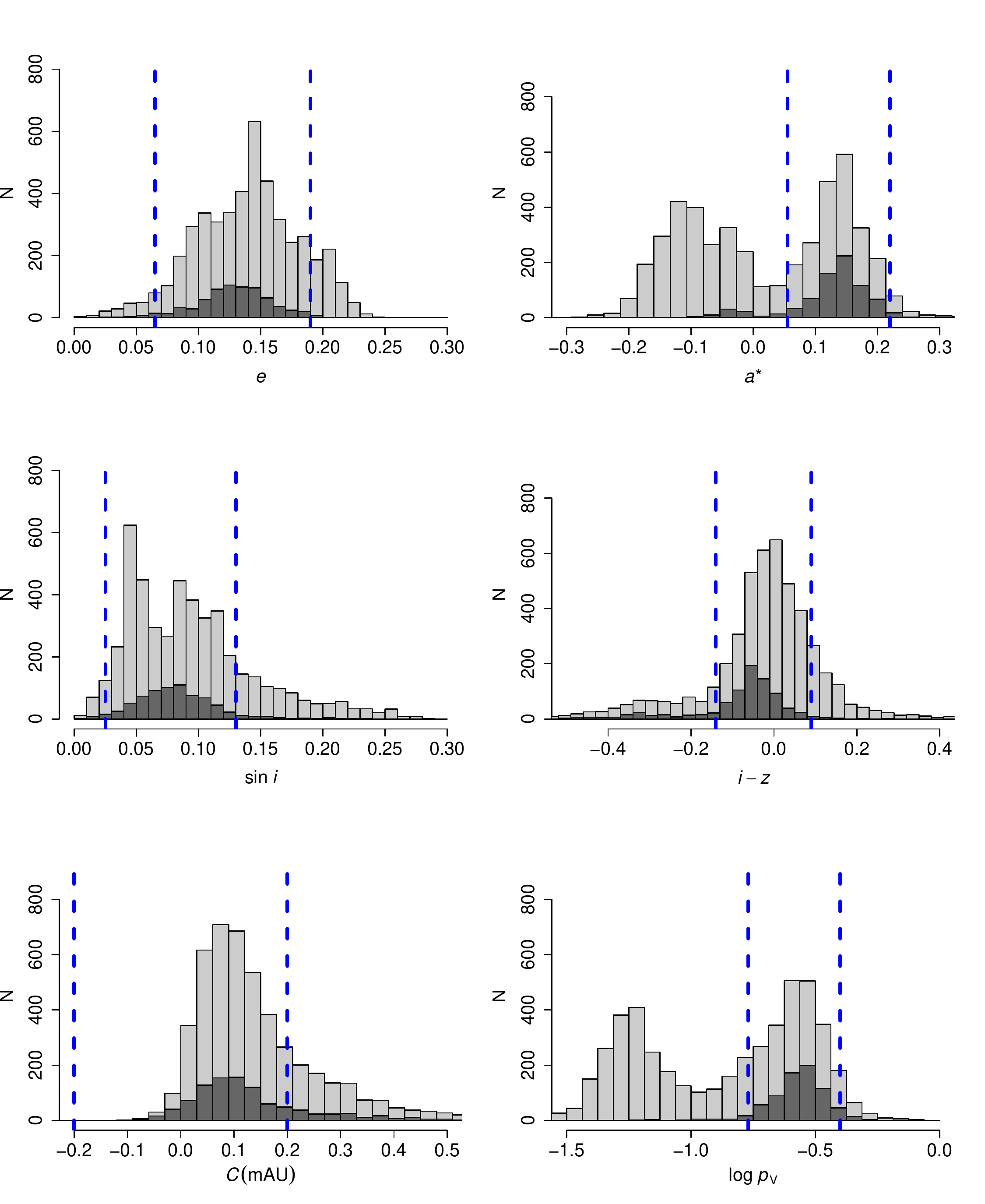}
\caption {\scriptsize Distributions of Flora-like objects (dark shading) against all 4696 background objects in Zone 1. Blue lines indicate the ranges selected in each dimension (see Table \ref{flora_properties}). For each parameter plotted, the dark shaded portions represent samples of all of the ``Flora-like'' objects that fall within the ranges in all other parameters, with the ranges specified in Table \ref{flora_properties}. For the $C$ parameter, a lower limit cannot be defined due to the removal of the retrograde-spinning fragments of the family via the $nu_6$ resonance; we have chosen a lower limit that would contain even those absent objects. Additionally, for the $C$ subplot no selection has been made in semimajor axis in order to show an unbiased distribution.
\label{09_cuts}}
\end{center}
\end{figure}

In practice, the process of determining the ranges is also an iterative one; each selection in one dimension increases the relative density of Floras in all of the other dimensions, allowing us to better define the ranges of the family in those other dimensions, which in turn allow us to better define the range for the original selection, etc. The strength of this iterative process is demonstrated in Figure \ref{09_cuts}. Each histogram shows the distribution of members from within the ranges of the Flora family against the distribution of all objects in Zone 1, as well as the ranges of our selection in that dimension. The restrictions in the other dimensions noticeably tighten the Flora distribution in each dimension. The ranges reported in the last column of Table \ref{flora_properties} represent the final step of the iteration for this family.

The 691 objects that have orbital and reflectance properties within these ranges (list and characteristics of these objects provided in online material, for now located in a CSV file at http://www.lpl.arizona.edu/$\sim$dykhuis/Flora-691.csv) represent a high-purity sample of Floras, which can serve as a high-priority target list for spectroscopic surveys attempting to probe the mineralogic properties of the Flora family. 

We note that the distribution of the family does not \emph{uniformly} fill these ranges; i.e., the ranges might be better described by gaussians of a given width rather than simple boundaries in each dimension \citep[e.g.,][see also discussion in Appendix \ref{appendixA}]{parker2008}. However, the multidimensional space used reduces the background sufficiently that each histogram in Figure \ref{09_cuts} shows a clear enhancement over background beginning at the parameter boundaries we select. That is to say, the distribution of Floras and of the background are distinctive enough that our description of the edges of the Flora dynamical zone are unambiguous. Future modeling of these results can be tested against both the observed extent of phase space and the variation of density with position in phase space. In the next section, we obtain a conservative estimate of the contamination due to interlopers in the sample.

\subsection{Estimate interlopers in the sample}
\label{interlopers}

\begin{figure}
\begin{center}
\includegraphics [width=6in]{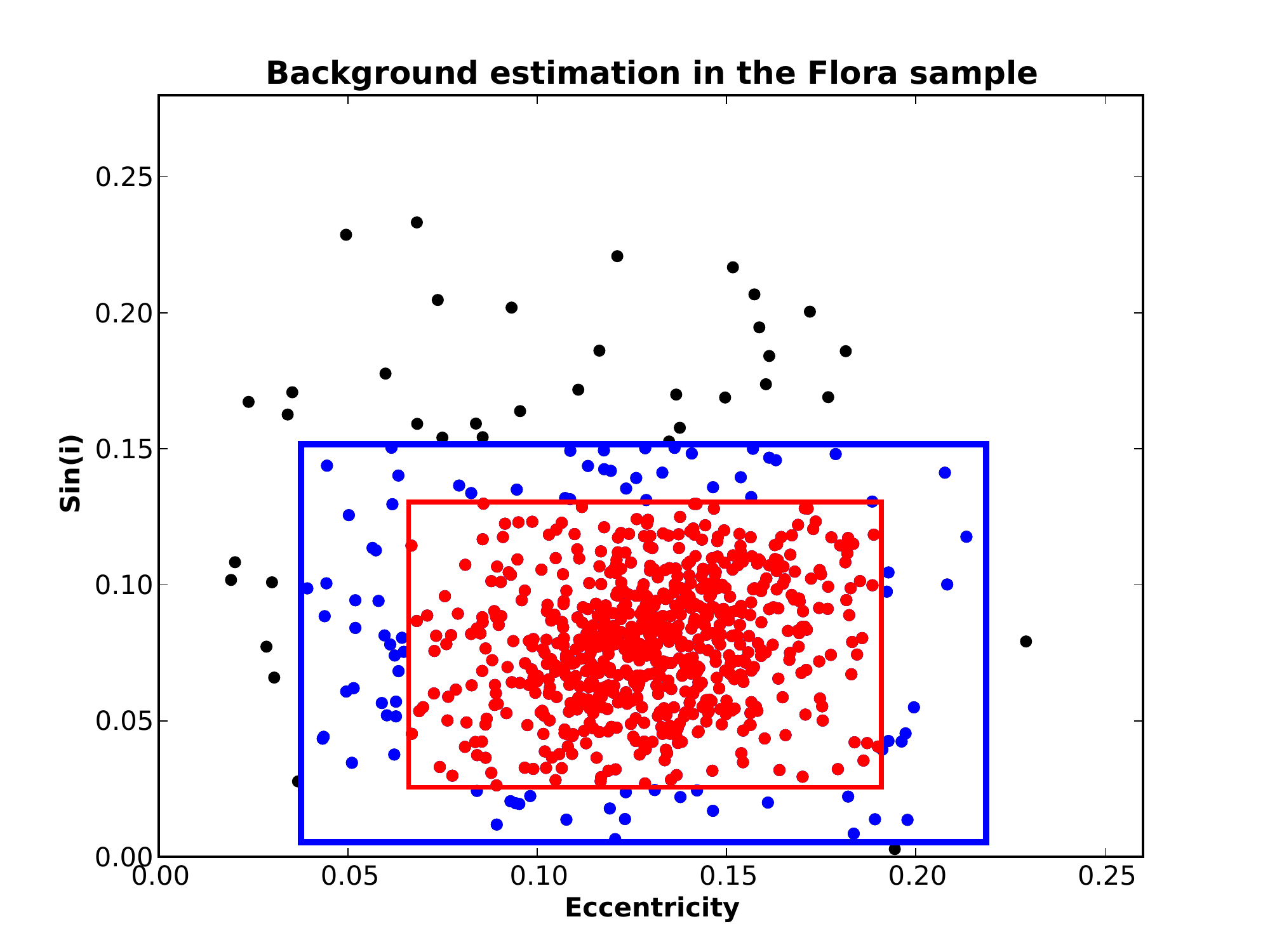}
\caption {Estimation of contamination from interlopers in our Flora sample. The red objects are the 691 that meet the criteria in Table \ref{flora_properties}; they define our Flora sample. The 84 blue objects lie within a nearby box of equal volume just outside the Flora sample range; we take these to be representative of our background density. We realize that dynamical dispersion in $e$ and $\sin i$ will place an unknown number of Flora family members in this phase space, however, the density of objects in the blue region allows us to place a conservative upper limit on the density of interlopers in the Flora sample. We estimate from this that the Flora sample contains an upper limit of 84 interlopers, or 12\% of the sample. 
\label{10_bkgd}}
\end{center}
\end{figure}

The above procedure yields a sample of objects that minimizes the fraction of non-Floras. Estimating that fraction is challenging because the Flora family spans nearly the entire inner main belt, a region with a background density that is far from uniform. Our method of defining the family's range in multidimensional space, however, enables us to separate the Flora family from most of this background contamination. Here we obtain a conservative upper limit on the percentage of expected interlopers in our sample by inspection of the distribution in $e$ and $\sin i$ (Figure \ref{10_bkgd}); objects located outside $0.065 < e < 0.19$ and $0.025 < \sin i < 0.13$ are less likely to be family members and provide an estimate of the number density of the background in the Flora region. We realize that objects with orbits beyond these ranges could be Flora members that have been dispersed into their current orbits via interactions with small resonances with Mars \citep[as described in][]{nesvorny2002}; however, the density distribution of objects beyond the sample range can provide a conservative upper limit on the expected density of interlopers in the sample. 

The Flora sample contains 691 objects; a sample from the region just beyond the family with the same volume in phase space contains 84 objects. We thus estimate that our Flora sample could contain up to 84 interlopers, or 12\% of the sample. 

As a further note that the Flora family benefits tremendously from the addition of reflectance information from the SDSS and WISE catalogs, we compute the interlopers in a sample of objects that is only defined by the dynamical ranges in Table \ref{flora_properties}. There are 1844 objects with color and albedo information within those dynamical ranges. Of those 1844 objects, 1074 or 58\% have SDSS colors that identify them as interlopers within the family. Of the same 1844 objects, 792 or 43\% have WISE albedos that identify them as interlopers. The color and albedo information is thus shown to be essential in separating the Flora family from the background families.

\section{Age Determination: Yarkovsky Drift in Semimajor Axis}
\label{yarkovsky}

Having identified the ranges of the Flora family in orbital and reflectance properties, we next evaluate the age of the family as determined by observations of the size-dependent Yarkovsky drift in semimajor axis. 

\begin{figure}
\begin{center}
\includegraphics [width=6in]{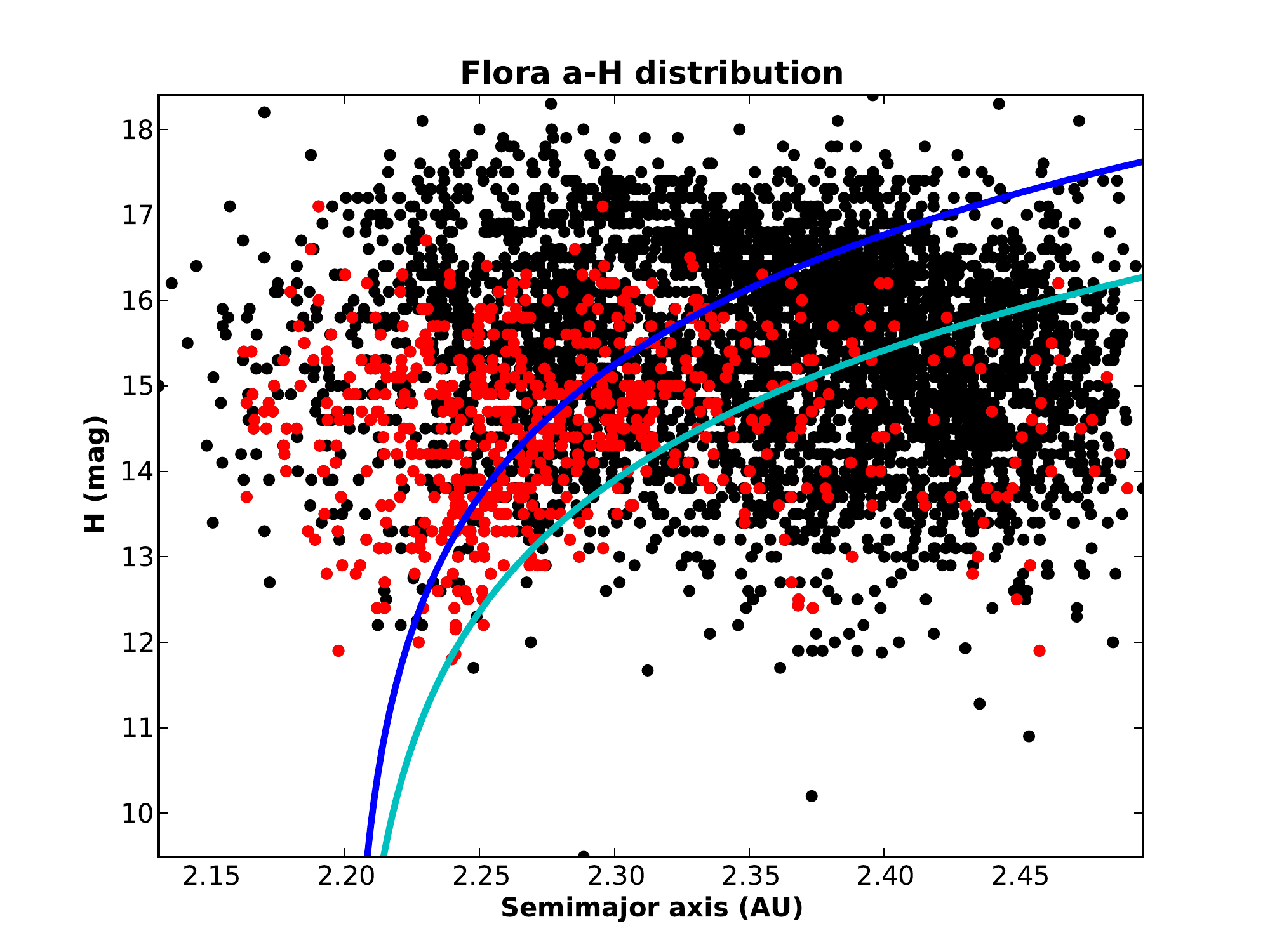}
\caption {Flora $a$-H distribution. Black objects are all 4696 Zone 1 objects observed by SDSS and WISE. Overplotted red objects are those that are likely Floras based on selections in $a^*$, $i-z$ and $p_{\text{V}}$, and also $e$ and $\sin i$; selection ranges are specified in Table \ref{flora_properties}. The ranges in $e$ and $\sin i$ have been tightened slightly, to $0.055 < e < 0.16$ and $0.05 < \sin i < 0.13$, in order to avoid contamination from the Campbell and Hertha families. The red objects have, on average, lower H values because H depends on albedo; in $a$-D space, the red and black populations have no vertical offset. The blue line represents the median $C$ value for the family, $C = 0.088$ mAU; the cyan line represents the outer edge of the family at $C = 0.164$ mAU.
\label{11_aH}}
\end{center}
\end{figure}

\begin{figure}
\begin{center}
\includegraphics [width=6in]{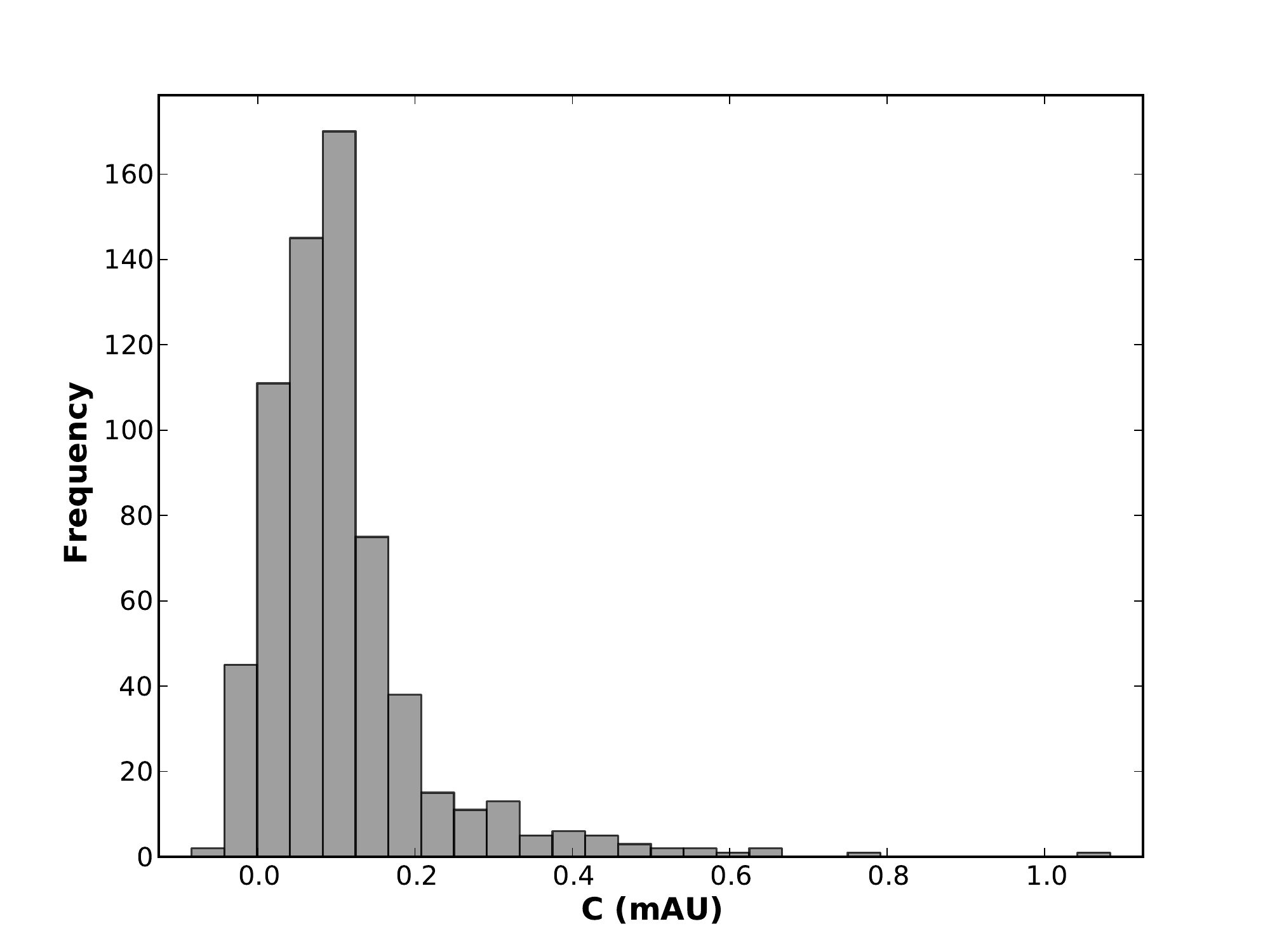}
\caption {Flora $C$ parameter distribution. Objects plotted are those that are likely Floras within selected ranges in $a^*$, $i-z$ and $p_{\text{V}}$, and also $e$ and $\sin i$; ranges are specified in Table \ref{flora_properties}. The ranges in $e$ and $\sin i$ have been tightened slightly, to $0.055 < e < 0.16$ and $0.05 < \sin i < 0.13$, in order to avoid contamination from the Campbell and Hertha families. The median $C$ parameter value is $0.088 \pm 0.002$ mAU, and the 1$\sigma$ value for $C$ is $C = 0.164$ mAU (fitting a gaussian to the right side of the distribution, this is the upper value of the range which would contain 68\% of the family). Using Equation \ref{yarcora3}, we calculate the age of the family to be $910^{+160}_{-120}$ My, using the $c_Y$ value determined for the S-type Karin family. The error is dominated by the uncertainty in $c_Y$.
\label{12_Cparam}}
\end{center}
\end{figure}

The plot of absolute magnitude vs. semimajor axis for this refined population shows half of the characteristic ``V'' shape that results from the Yarkovsky spreading around the original collision (Figure \ref{11_aH}). The distribution in $C$ parameter, calculated for each member of the population via Equation \ref{Cparam}, is shown in Figure \ref{12_Cparam}. The median value of $C$ for the prograde Floras is $0.088 \pm 0.002$ mAU, and the value of $C$ that best defines the outer edge of the family is $C = 0.164$ mAU. (If we fit a gaussian to the right side of the Flora peak in Figure \ref{12_Cparam}, this represents the 1$\sigma$ value for $C$, the upper value of the range which would contain 68\% of the family. This agrees with a cursory visual inspection of the distribution in Figure \ref{11_aH}, and thus we take $C = 0.164$ mAU to be a reasonable outer limit for the family.) 

It is possible to extract the age of the family from the $C$ parameter distribution via the relation (Molnar et al. in prep):

\begin{equation}
\label{yarcora3}
t = \frac{1329 \cdot C \cdot {a_\text{P}}^2}{c_Y \sqrt{p_\text{V}}(1-p_\text{V})\cos\epsilon}
\end{equation}

\noindent where $t$ is the time (in My) since the collision, $a_P$ is the semimajor axis of the parent body (in AU), in this case 8 Flora, $\epsilon$ is the obliquity, and $C$ is given in AU. The coefficient $c_Y$ is a complex function of the asteroid's material properties: thermal conductivity, specific heat, and material density \citep[for a discussion of the dependence of the Yarkovsky drift on these parameters, see, e.g.,][]{bottke2006}.

These material properties are unknown and difficult to estimate from first principles \citep{cellino2009}, and thus introduce considerable uncertainty into the value of $c_Y$ appropriate to the Flora family. Here we instead assume that the $c_Y$ value for the Flora family is the same as that of another S-type family, the Karin family, for which we have a collisional age determined from direct backward integration of the members' orbital elements \citep[e.g.,][]{nesvorny2004}. This 5.8-My-old family exhibits a Yarkovsky spreading consistent with a $c_Y$ parameter in the range of $0.0026 \text{ AU$^3$km/My} < c_Y < 0.0035$ AU$^3$km/My (Molnar et al. in prep). This corresponds to a maximum Yarkovsky drift rate in the range of 2.9-3.9 x $10^{-4}$ AU/My (44-59 m/yr) for a 1-km object with an albedo of 0.3 at $a = 2.5$ AU, which is consistent with typical drift rates measured for near-Earth objects \citep[20-70 x $10^{-4}$ AU/My,][]{nugent2012}, when adjusted for semimajor axis offset. We adopt the range of $c_Y$ found for the Karin family in our calculation of the age of the Flora family. 

The obliquity dependence of the Yarkovsky drift (Equation \ref{yarcora3}) introduces an important factor in the age determination. Assuming that the highest values of $C$ in the population correspond to prograde objects with spin axes perpendicular to their orbital plane ($\cos\epsilon$ = 1), calculation of the age via Equation \ref{yarcora3} is straightforward. Taking the outer edge of the family to be $C = 0.164$ mAU, the average albedo $p_\text{V}$ to be 0.291, the diameter $D(H,\langle p_\text{V}\rangle=0.291)$ as given in Equation \ref{pV_eq}, the semimajor axis to be $a_\text{F} = 2.201$, and the value for $c_Y$ to be 0.00305 AU$^3$km/My, we find an age for the Flora family as determined by the drift observed in Figure \ref{11_aH} of $910^{+160}_{-120}$ My. The error in this estimate is dominated by the uncertainty in $c_Y$.

However, any small asteroid is likely to have had its obliquity modified over the lifetime of the family, for example by non-destructive impact events or radiative torques such as the Yarkovsky-O'Keefe-Radzievskii-Paddack (YORP) effect \citep[e.g.,][and references therein]{bottke2006}. In addition, spin-orbit resonances are also known to affect the obliquity distributions of asteroids, including for example the resonant obliquity state known to affect the asteroid 951 Gaspra (whose orbital and albedo parameters are consistent with Flora membership, but is not included in our selection because its SDSS color parameters are unknown). \citet{rubincam2002} find that 951 Gaspra's spin precession rate and orbital precession rate are nearly commensurate, forcing large variations in the asteroid's obliquity over short (My) timescales, and they speculate that the YORP effect drove 951 Gaspra into this spin-orbit resonance. In a later work, \citet{vokrouhlicky2006b} studied the evolution of five asteroids (including Gaspra) spaced throughout the main belt, and confirmed that overlapping spin-orbit resonances can force large obliquity variations. 

If a significant fraction of the Flora family members in our sample have been affected by spin-orbit resonances, the median value of $C$ = 0.088 mAU (cf. Figure \ref{12_Cparam}) would be more representative of the drift than the maximum value $C$ = 0.164 mAU. Since the timescales for obliquity variation are much, much shorter than the likely age of the family, the relevant obliquity would be the time average obliquity of the family members ($\cos\epsilon = 0.5$). Using these values in Equation \ref{yarcora3} yields nearly the same age as above, because the change in age resulting from using the median value of $C$ (rather than the extreme) is largely canceled by the simultaneous use of the lower value of $\cos\epsilon$. Because of the actual processes affecting obliquity histories remain a subject of much debate in the field (see discussion in Section \ref{discussion} below), we adopt for now the result based on maximum $C$ and constant obliquity: $910^{+160}_{-120}$ My. Future study of the distribution of the family as a function of both H and $C$ may help to resolve the debate.

\section{Discussion}
\label{discussion}

Our approach to identifying Flora family members has refined understanding of the median orbital and reflectance properties of the family, as well as the family's ranges in the parameter space (Table \ref{flora_properties}). The determination of these ranges provides an opportunity for improved modeling and interpretation of dynamical evolution. The addition of color and albedo information gives us a tremendous advantage in disentangling the overlapping families in the inner main belt; therefore we find it essential to include this information in our analysis of the Flora family, although we recognize the incompleteness of those catalogs leads to incompleteness in our samples. However, since the Flora family is numerous, we are still able to explore the full range of dynamical phase space which it occupies, which enables us to extract physical information about the family. (While our analysis is a first step toward obtaining the size-frequency distribution of the family, the incompleteness of our sample does prevent us from drawing conclusions about the size-frequency distribution at this time, see Section \ref{conclusion} for further discussion on this.)

\subsection{Discussion of Flora Family Age}
\label{age_implications}

The age of the Flora family derived here (790-1070 My) is consistent with some previous estimates, and somewhat older than others. \citet{nesvorny2002} found ages from consideration of the family's dispersion in $e$ and $\sin i$ under the influence of small resonances with Mars: dispersion in $e$ gave an age of 900 My, dispersion in $\sin i$ yielded an age of 500 My. Given the approximations involved in their method, they considered these two numbers to be roughly in agreement, and concluded the age is likely $<$ 1 Gy. In a later work \citep{nesvorny2007}, they connected the 500-My age with a major impact event around 470 Mya evident among the L chondrite meteorites; however, our age result rules out this connection. \citet{hanus2013} obtained an age of 1000 $\pm$ 500 My via comparison between observed and simulated distributions of family member obliquities correlated with proper semimajor axis, a result consistent with ours, but with much wider uncertainty. 

Crater populations on 951 Gaspra have produced a wide range of estimates for the age of its surface: from 50 My \citep{greenberg1994}; 65-100 My \citep{obrien2006}; 200 My \citep{chapman1996}; and 1600-3000 My \citep{marchi2013}. However, the surface age is not necessarily the same as the age of the asteroid itself; only the latter solution is inconsistent with our results. Note also that surface age estimates depend on assumptions about the impact flux of very small bodies.

The largest sources of uncertainty in our age estimate come from 1) the definition of the outer edge of the family in $a$-H space, specified by the $C$ parameter in Equation \ref{yarcora3}, 2) the dynamical model used to approximate family member obliquities (possibly influenced by spin-orbit resonances and YORP), and 3) the assumption of similar material properties --- and thus $c_Y$ --- between the Flora and Karin families (especially densities).

\begin{figure}
\begin{center}
\includegraphics [width=2.6in]{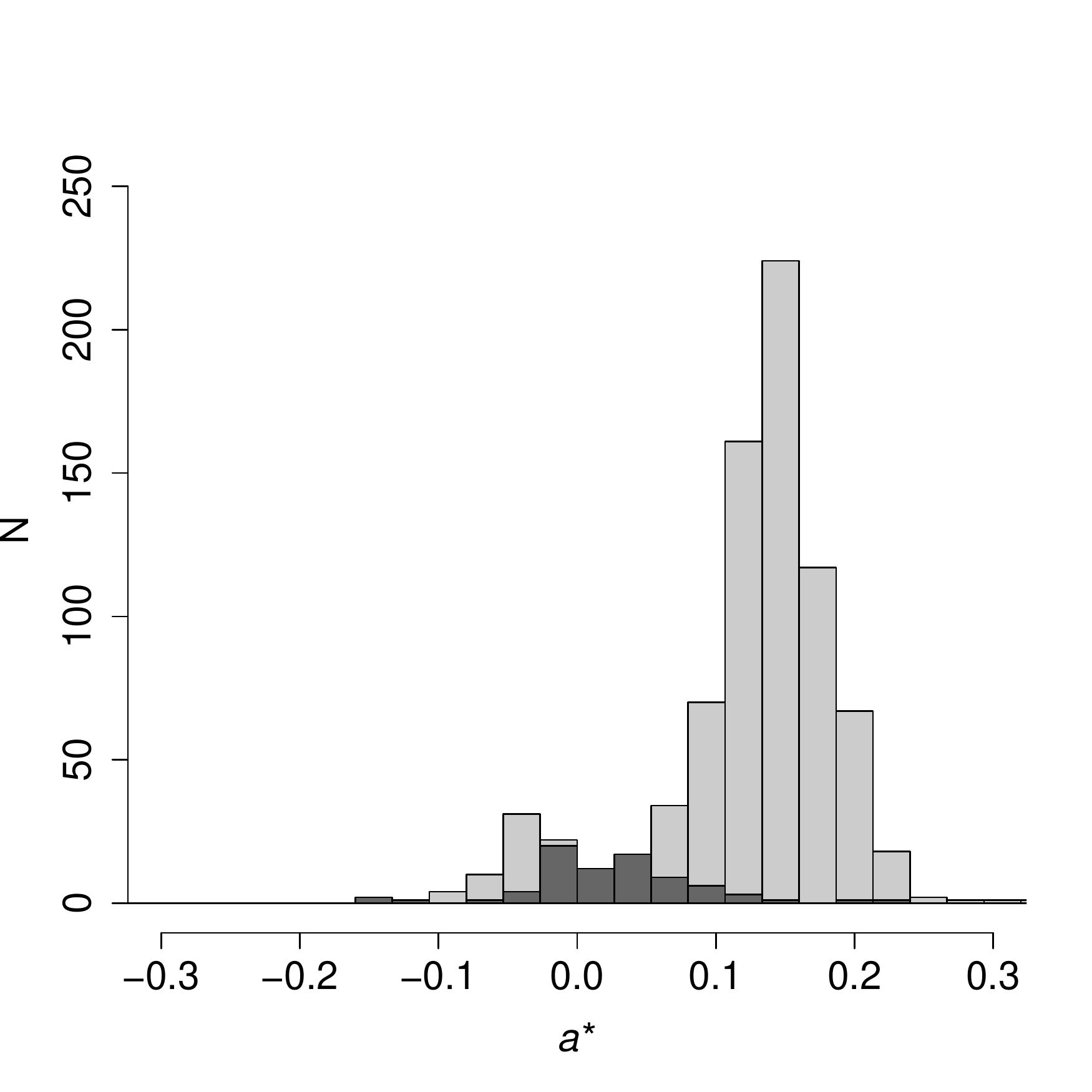}
\includegraphics [width=2.6in]{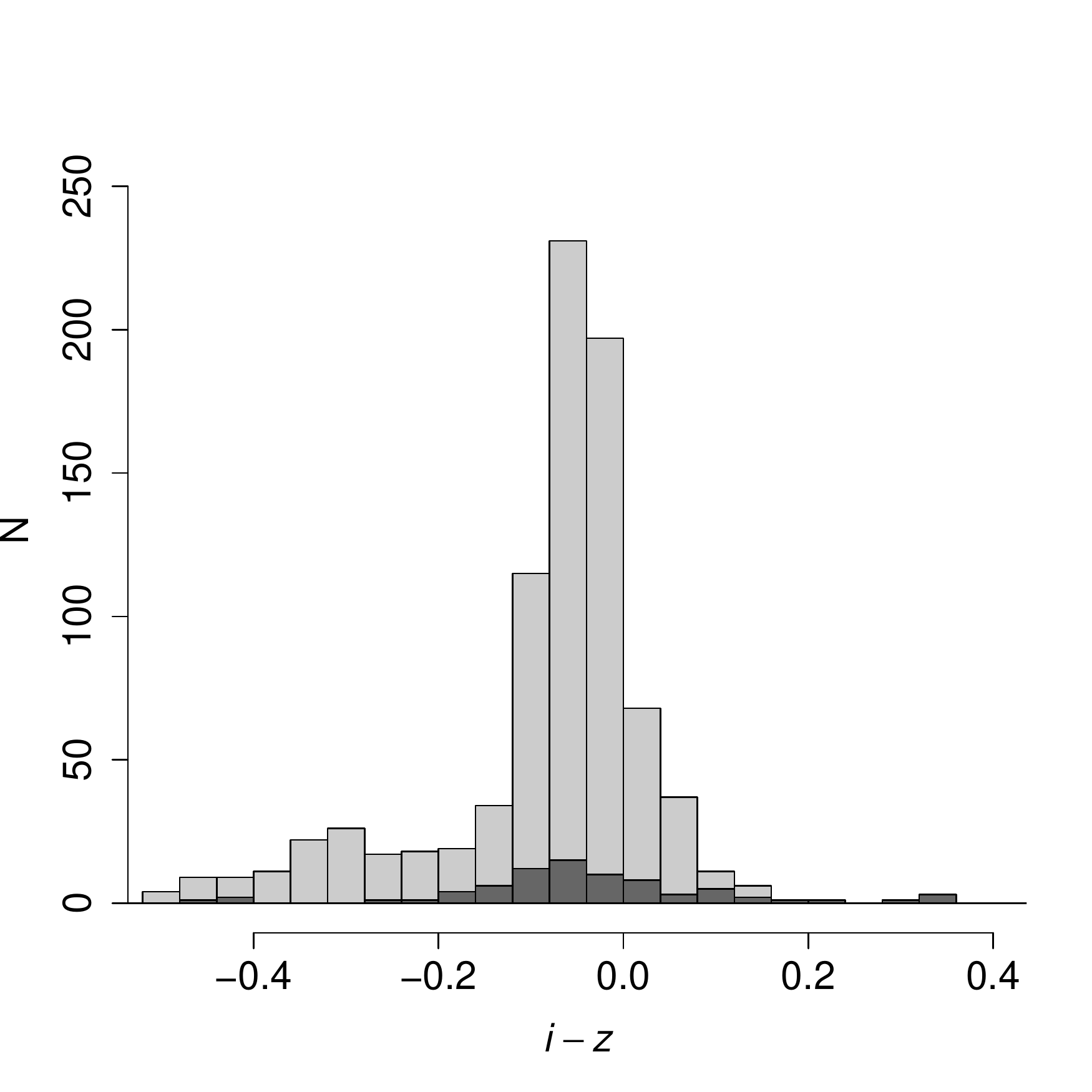}
\includegraphics [width=2.6in]{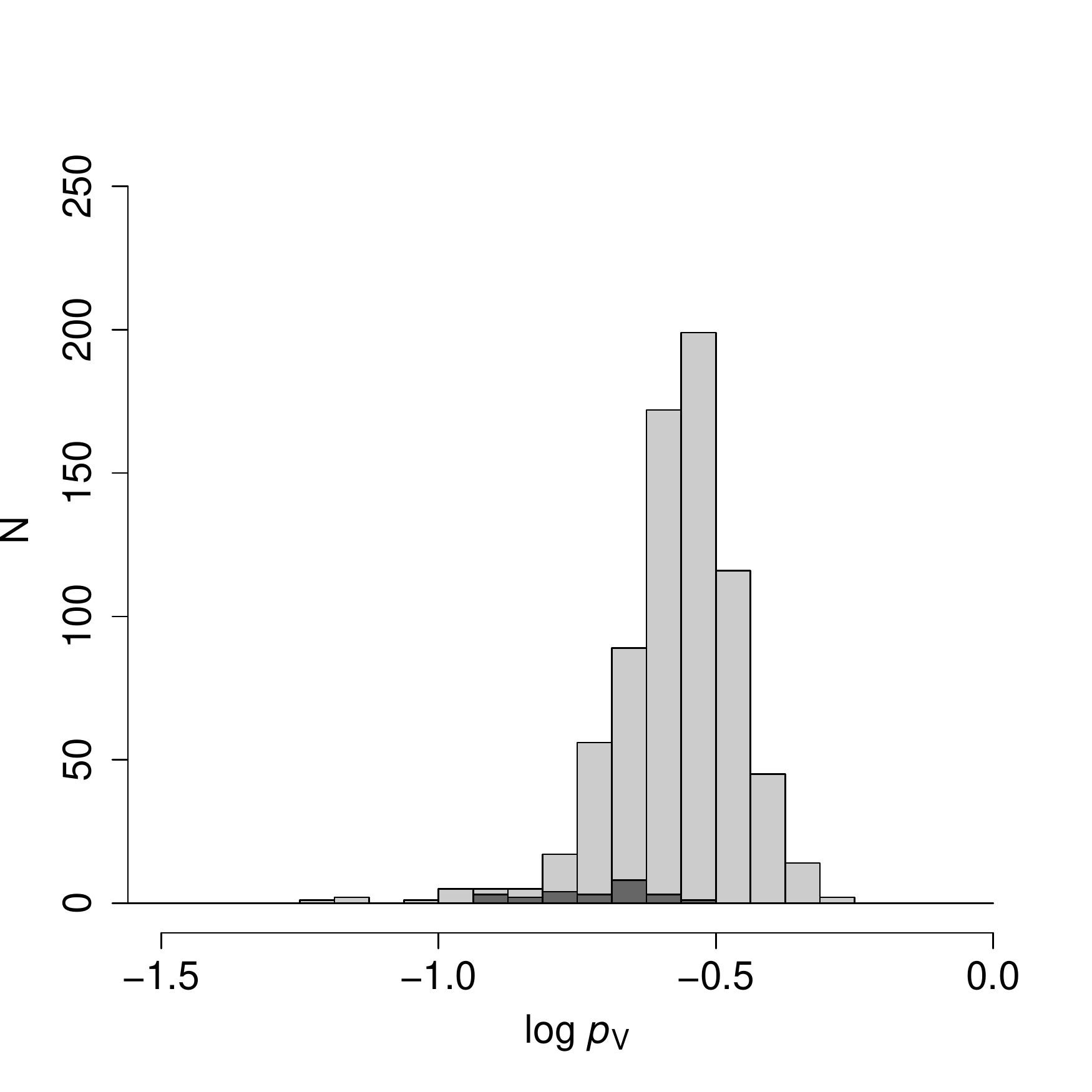}
\caption {Comparison of Flora family (light shading) and Karin family (dark shading) reflectance properties. The young Karin family has median reflectance properties of $a^* = 0.023 \pm 0.013$, $i-z = -0.052 \pm 0.015$, and $p_\text{V} = 0.206 \pm 0.018$. We expect the median $a^*$ color to shift slightly toward higher values with age due to space weathering processes; e.g., the 2-By-old Koronis family shows an average $a^*$ color of 0.09, much higher than that of the Karin family. The Flora family has a median value of $a^*$ (0.126 $\pm$ 0.007) that is higher than that expected for a family of similar age among the Koronis families, even with the effects of space weathering taken into account (see Figure \ref{14_space_weathering}).
\label{13_FloraKarin}}
\end{center}
\end{figure}

\begin{figure}
\begin{center}
\includegraphics [width=6in]{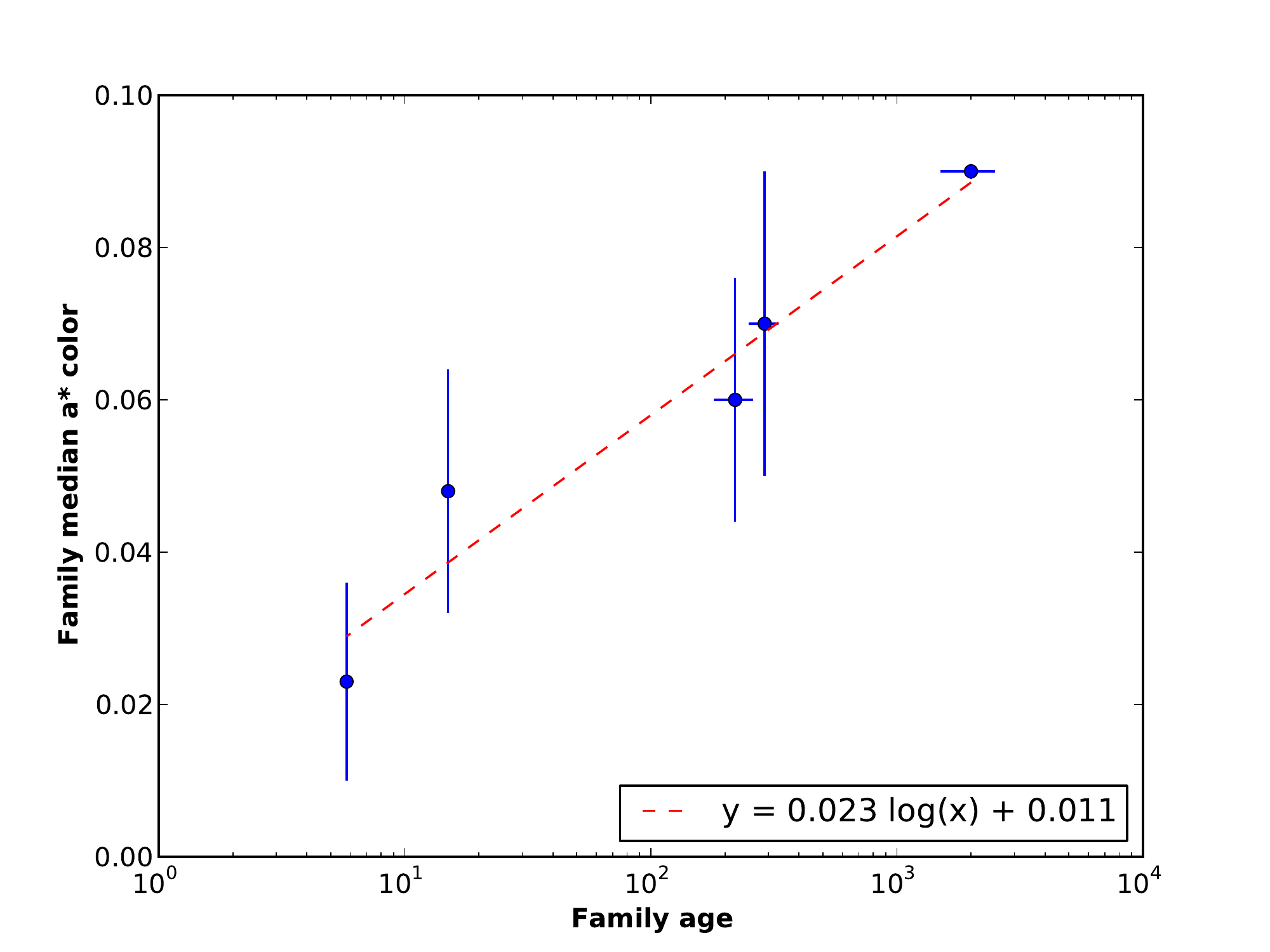}
\caption {From \citet{molnar2011}; space weathering trend in $a^*$ color visible among four subfamilies of the 2-By-old Koronis family, associated with 832 Karin (5.8 My), 462 Eriphyla (220 My), 321 Florentina (290 My), and a small cratering event on 158 Koronis itself (15 My). The $a^*$ color increases slightly from 0.023 to 0.09 in about two billion years among the S-type objects of this family. Similar plots for $i-z$ and $p_\text{V}$ show no clear trends.
\label{14_space_weathering}}
\end{center}
\end{figure}

In fact it is possible that there are differences in material properties between Flora and Karin family members. Figure \ref{13_FloraKarin} shows the colors and albedos of the Flora and Karin populations. The differences in $a^*$ between the two families can partly be accounted for by space weathering, which has introduced systematic changes in $a^*$ among four subfamilies of the Koronis family, one of which is Karin (see Figure \ref{14_space_weathering}, after \citet{molnar2011}). However the differences in $a^*$ color and $p_\text{V}$ do suggest some additional differences at least in surface properties. These differences could result in changes in the Yarkovsky parameter $c_Y$.

Differences in bulk densities between the two families would certainly affect the $c_Y$ parameter. Unfortunately, due to the paucity of density data, there are very few density measurements for the members in our Flora or Karin samples. In a recent review by \citet{carry2012}, the S-type asteroids as a group are stated to have average densities of 2.7 g/$\text{cm}^3$, while individual asteroids 8 Flora and 42 Ariadne are stated to have densities of 6.50 $\pm$ 1.28 and 8.99 $\pm$ 2.57 g/$\text{cm}^3$ respectively. These values seem significantly higher than the average for S-type objects, and are in disagreement with the density obtained for 8 Flora by \citet{konopliv2011} (1.84 g/$\text{cm}^3$). Another object, 1089 Tama (which falls within our Flora sample ranges in orbital elements, but lacks color and albedo information) has a density of 2.52 $\pm$ 0.29 g/$\text{cm}^3$ \citep{carry2012}. Additionally, density information will soon be available for binary asteroid 939 Isberga (which falls within our Flora sample ranges in orbital elements and albedo, but lacks SDSS color information). Detailed volume information exists for asteroid 951 Gaspra; however, the Galileo spacecraft did not fly close enough to Gaspra to obtain a reliable mass/density estimate.

For the Karin family densities, we turn to the larger Koronis family within which the Karins are a subfamily. The two Koronis family asteroids with density data are 243 Ida and 720 Bohlinia; their densities are 2.35 $\pm$ 0.29 and 2.74 $\pm$ 0.56 g/$\text{cm}^3$, respectively \citep{carry2012}. These values are consistent with S-type densities, and are not inconsistent with the Flora densities, given the large uncertainties. Additional density data will be needed to better constrain our assumption of similar $c_Y$ values between these two families.

The comparison between the Karin and Flora families is also only technically valid for objects with 11 mag $< \text{H} <$ 17.5 mag (1 km $< \text{D} < $4 km), the size range of the objects in the Karin family. This could, in principle, affect our $a$-H analysis of the Flora family age; however, a repeat of the analysis in Section \ref{yarkovsky} for all objects in the Flora family with 11 mag $< \text{H} <$ 17.5 mag only affects the age estimate slightly, well within the uncertainty of our solution. 

Lastly, the age determination depends in part on assumptions about the obliquities of the Flora family asteroids, the latter of which are influenced by the YORP effect. The YORP torques can cause a rapid increase in an asteroid's spin rate until material is shed from the equator \citep{walsh2008}, or a rapid decrease in spin rate until a ``tumbling,'' or non-principal axis rotation, state is reached \citep{vokrouhlicky2002,bottke2006}. Small changes in topography and surface geometry can affect the strength of these YORP torques, possibly turning the process of moving to the end-states into a random walk \citep{nesvorny2008,statler2009,cotto-figueroa2013}. The effects of this ``variable YORP'' \citep{bottke2013}, as well as a quantification of the tendency of family members to become caught in spin-orbit resonances, will be the topic of our future work; however, preliminary analysis (discussed at the end of Section \ref{yarkovsky}) suggests that both spin-orbit and variable YORP effects will not significantly affect the age. Our detailed study of the current spin states of the objects within our Flora sample follows in Section \ref{spins}.

\subsection{Spectra of Flora Family Objects}
\label{spectra}

The dynamical overlaps and complexity of the Flora region has led to the miscategorization of background family members and miscellaneous objects as Floras, and vice versa, confusing the question of Flora member taxonomy and underlying mineralogy. \citet{nesvorny2005} proposed that the Flora family itself could be taxonomically mixed. Our analysis in Section \ref{reflectance} demonstrates that the Floras do have a characteristic range of SDSS color and WISE albedo that sets them apart from the surrounding families, and that the shape of their distribution in orbital element space is consistent with a single collisional family. Taxonomic inhomogeneities, which would manifest themselves as two or more families with similar orbits but different reflectance properties, are not seen.

\begin{figure}
\begin{center}
\includegraphics [width=6in]{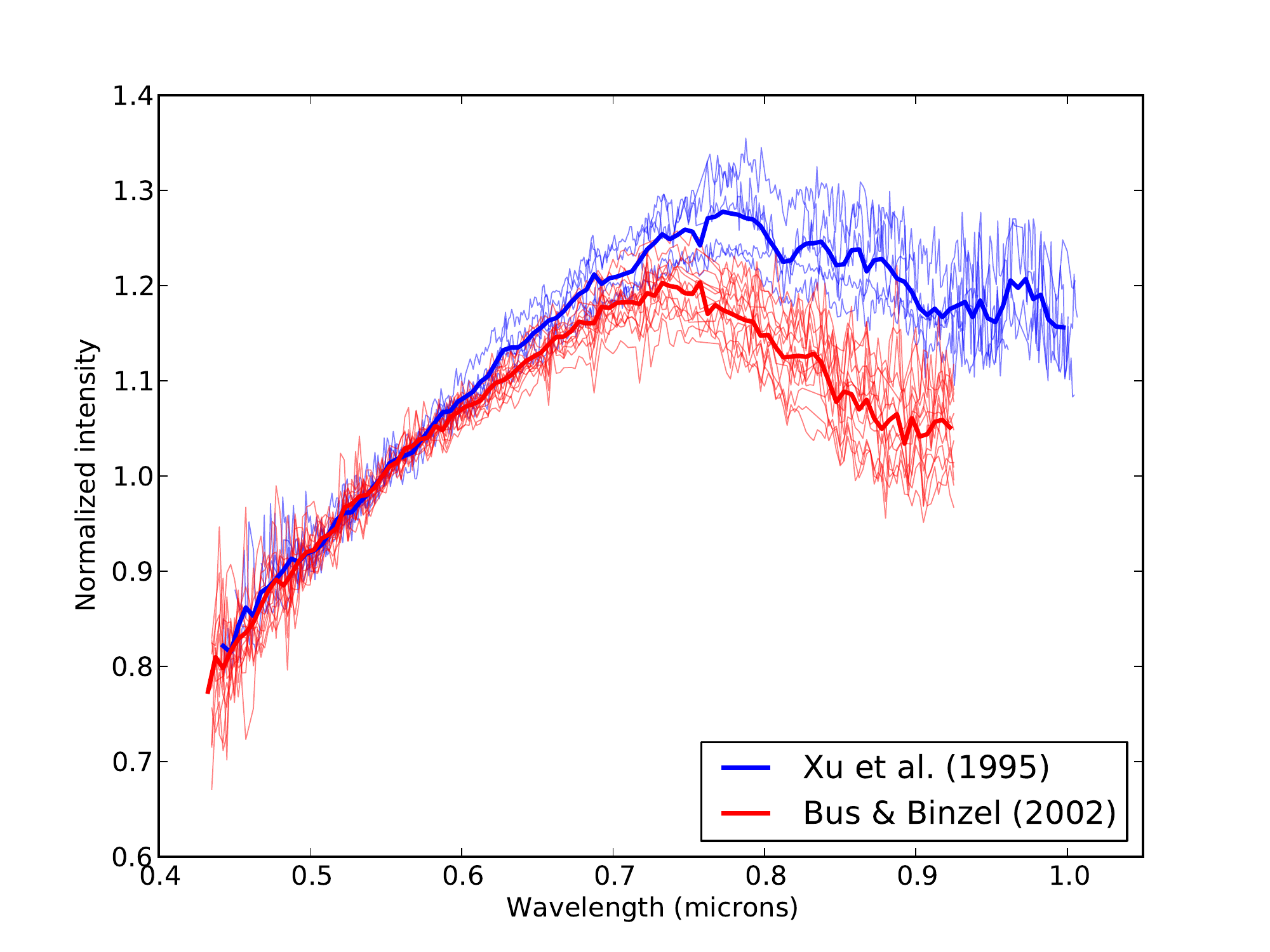}
\caption {Spectra of 19 objects from our Flora core group. The blue lines are 5 objects that were observed by SMASS I \citep{xu1995}, and the red lines are spectra of 14 objects that were observed by SMASS II \citep{bus2002}. The bold lines are running means for the two sets. The differences in slope near 0.8 microns are due to systematic differences between the two surveys. The Flora objects show the characteristic features typical of S-type objects.
\label{15_spectra}}
\end{center}
\end{figure}

Spectral data are available from the SMASS online database\footnote{smass.mit.edu, accessed March 3, 2014} for 19 objects from our sample (with fewer than four of these objects expected to be interlopers, Section \ref{interlopers}). Spectra are available in the literature for an additional 12 objects, but are not formatted for easy comparison here. The spectra of the 19 (specifically objects numbered 8, 819, 1798, 2019, 2119, 2410, 2467, 3121, 3181, 3658, 3677, 3841, 3972, 4001, 4025, 4287, 4640, 4650, and 5008) are plotted in Figure \ref{15_spectra}. When the systematic difference between surveys are accounted for, the spectra look very similar, and all show the characteristic features typical of S-type objects. 

Due to the correlation of spectra with SDSS color, the similarity of these spectra is to be expected. Thus the similarity of these spectra does not in itself require a homogeneous underlying mineralogy of the Flora collisional family; rather, the homogeneity of the family is demonstrated in the homogeneity of the SDSS colors and WISE albedos in the dynamical core of the family. The spectra, however, do provide a more detailed probe into the mineralogy of some of the objects in the Flora sample.


\subsection{Spin Distribution of Flora Family Objects}
\label{spins}

\begin{figure}
\begin{center}
\includegraphics [width=6in]{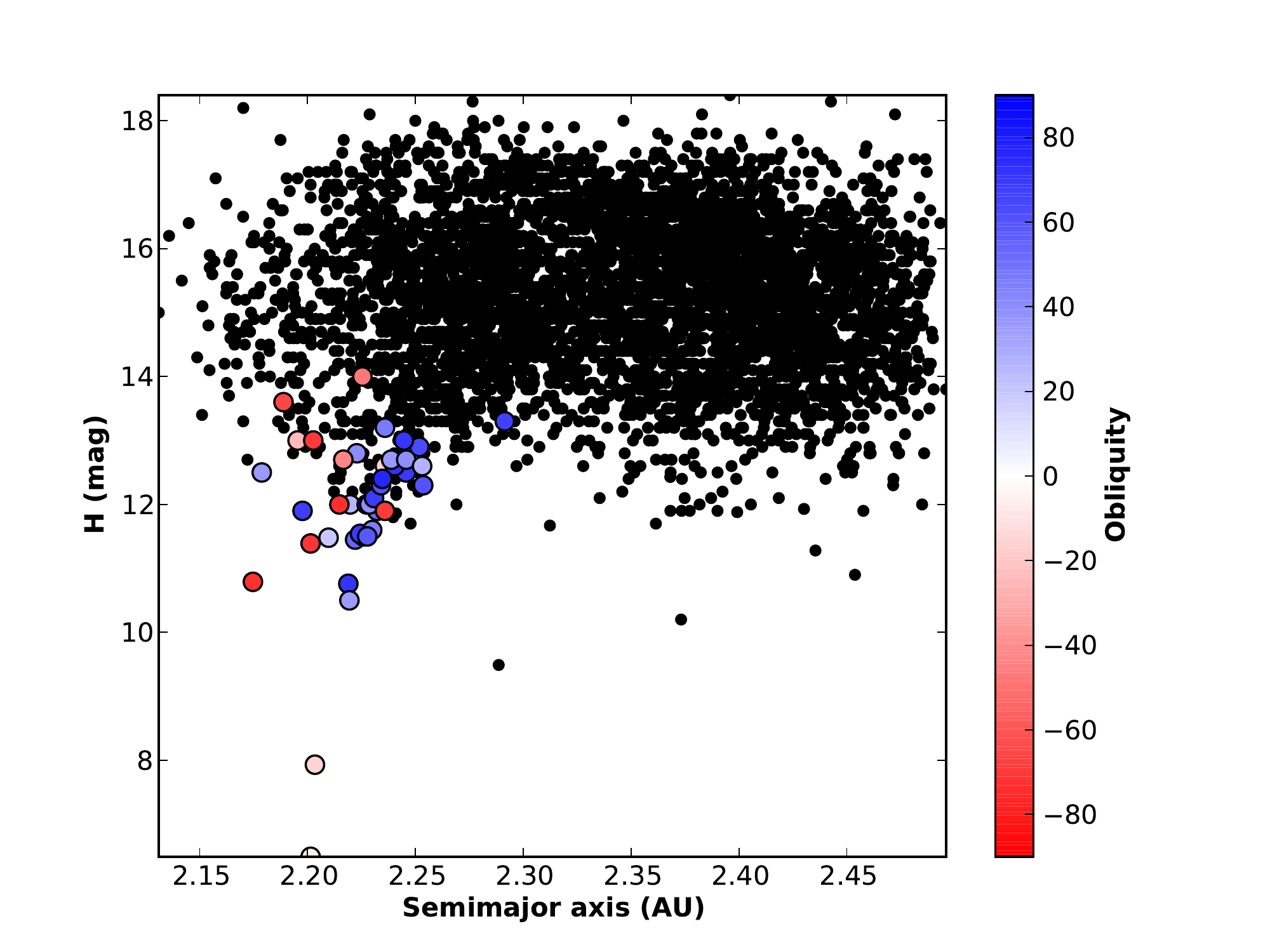}
\caption {Distribution of Flora family member obliquities. The abundance of prograde rotators near $a$ = 2.23 AU, H = 12 magnitudes is consistent with the Yarkovsky effect dispersing the remnants of the original collision that formed the Flora family.
\label{16_obliquity}}
\end{center}
\end{figure}

\begin{figure}
\begin{center}
\includegraphics [width=6in]{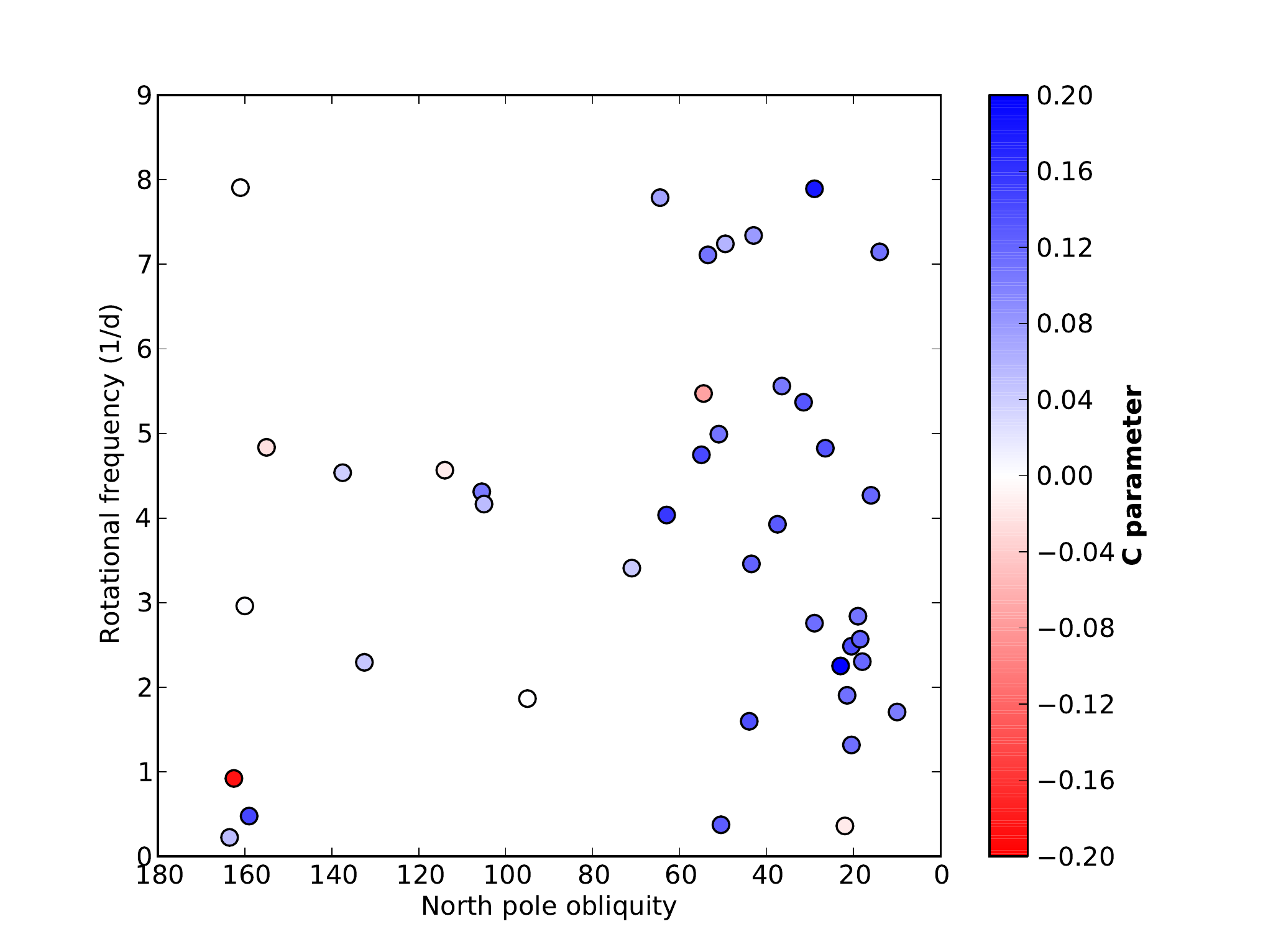}
\caption {Distribution of Flora obliquities and rotation rates for the 41 objects in our Flora sample with spin information reported in the DAMIT database (cf. \citet{kryszczynska2013} Figure 2). Colors represent the extent of drift in semimajor axis from the family center, the $C$ parameter; blue objects are those that have drifted to higher semimajor axes. We find no evidence of Slivan states among the Flora family asteroids. 
\label{17_ak2}}
\end{center}
\end{figure}

The spin axis distribution of the objects in our sample of Flora family members can in theory differentiate between dynamical models for the family's evolution. The obliquity distribution for the 41 objects in our sample of Floras which have obliquities reported in the DAMIT database\footnote{http://astro.troja.mff.cuni.cz/projects/asteroids3D/, accessed February 19, 2014} is shown in Figure \ref{16_obliquity}. We confirm an abundance of prograde rotators present in the high-semimajor axis wing of the family in $a$-H space, consistent with model predictions of the Yarkovsky drift and with earlier studies by \citet{haegert2009}, \citet{kryszczynska2013} and \citet{hanus2013}. The distribution of obliquities among the asteroids in our sample does not show clear evidence of the capture of the Flora family members into ``Slivan states'' \citep[Figure \ref{17_ak2}][]{slivan2003,vokrouhlicky2003,kryszczynska2013}. The particular Slivan state spin-orbit resonance is not expected to occur in this region of the proper element space. 

\begin{figure}
\begin{center}
\includegraphics [width=6in]{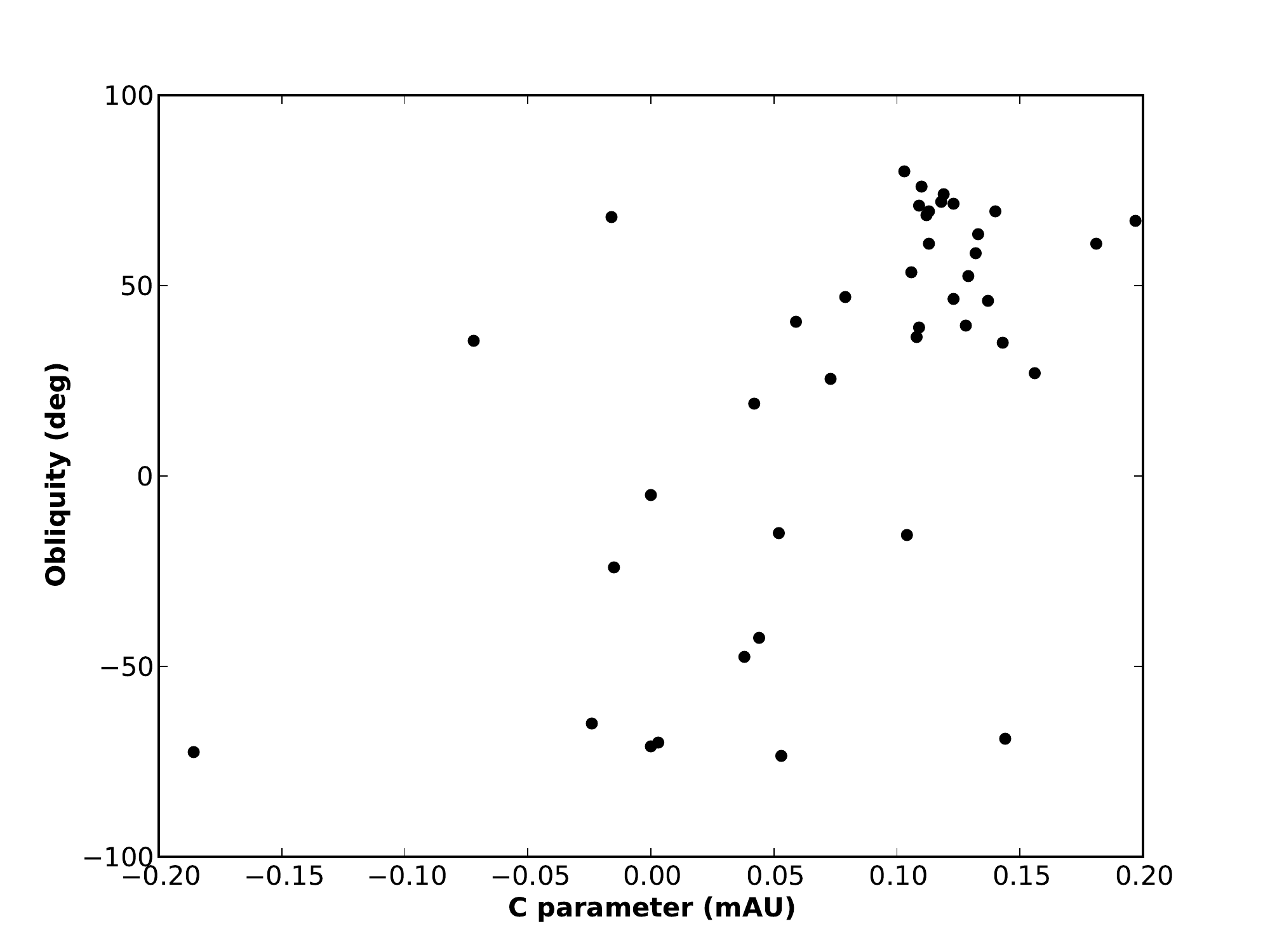}
\caption {Distribution of Flora sample obliquities for different $C$ parameters. If the members of the Flora family are affected by spin-orbit resonance, we would expect to see the objects near the median $C$ parameter display a wide range of obliquity values. We see here a cluster of objects between 0.1 mAU $< C <$ 0.15 mAU and 30 deg $< \beta <$ 80 deg. With the data available, we conclude that the Flora family shows a mild signature of spin-orbit resonance capture among its prograde members.
\label{18_Cbeta}}
\end{center}
\end{figure}

While Slivan states are not observed or expected, a significant fraction of the Floras could be caught in the same sort of spin-orbit resonance as the one that drives the obliquity variation of 951 Gaspra (see Section \ref{yarkovsky}). The observational signature of this resonance would be a cluster of both the prograde and retrograde objects around the semimajor axis values that correspond to the amount of drift expected for the time-average value of obliquity (the median $C$ parameter). If the members of the Flora family are indeed affected by this spin-orbit resonance, we would expect to see the objects near the median $C$ parameter (about $C = 0.088$ mAU among the larger sample of Floras, slightly higher than that in this smaller group) to have a wide range of obliquity values. If there were no ongoing change in obliquity, we would expect a strong correlation between obliquity and $C$. Figure \ref{18_Cbeta} shows no such correlation; rather, we observe a cluster of objects between 0.1 mAU $< C <$ 0.15 mAU and 30 deg $< \beta <$ 80 deg. This range of obliquity values is slightly tighter than the range of 10 deg $< \beta <$ 80 deg found for 951 Gaspra in the spin-orbit simulations done by \citet{vokrouhlicky2006b}. We conclude, therefore, that the Flora family shows a mild signature of spin-orbit resonance capture among its prograde members, but that more data are needed to make a strong claim in either direction.

\section{Conclusion}
\label{conclusion}

Current efforts in the literature to assign family membership in the inner main belt use algorithms such as the HCM to identify clusters from the background in multidimensional phase space \citep{nesvorny2012,carruba2013,milani2013}. These algorithms succeed in identifying clusters, but fall short of investigating the ranges of families in the phase space. Our approach uses the full multidimensional parameter space to reduce interlopers to a negligible fraction and hence allow the identification of trustworthy boundaries for the family. These boundaries contain important physical information about the family, information which is lost if the family is split in pieces or combined with unrelated families. 

\citet{milani2013} explicitly note that their automated approach to family identification does not eliminate the need for additional non-automated investigations; our work thus compliments the automated methods in providing an in-depth analysis of a particular family which defies investigation via the automated approach. For example, the automated algorithms of \citet{carruba2013} and \citet{milani2013} were both unable to identify the Flora family as a collisional or dynamical grouping, instead labeling the core of the larger, dispersed Flora family as the ``1052 Belgica family'' and the ``1338 Duponta family'', respectively. The signature of the Floras in $a$-H space rules out these objects as parents of the collisional family. 

While the ranges of the Flora family reported here are inclusive, the sample of objects reported is not the complete family membership. This distinction is important, because the extraction of collisional information from the size-frequency distribution (SFD) of a sample of family members assumes that the distribution of the sample is an accurate representation of the underlying distribution of the family. Our definition of the ranges of the Flora family in orbital and reflectance properties is a necessary first step toward obtaining the family's SFD, but it is not sufficient: in order to obtain the full SFD we would need to have full reflectance information for all of the objects within the Flora ranges in orbital proper elements, to obtain a fully inclusive sample. Our sample necessarily discards objects that are likely to be Floras based on dynamical considerations, but for whom reflectance data are not available. It must also be noted that family identification methods that neglect reflectance information such as color and albedo can be expected to produce family lists for the Flora family that contain at least 50\% interlopers, which likely alter the observed SFD from the true Flora SFD.

\subsection{Our Method: Key Points}

Our method also uses the following key techniques to characterize the Flora family:

\begin{itemize}
 \item We use an iterative procedure to identify the characteristic properties of the family by focusing in on the ``core'' of the family in multidimensional space. This allows us to largely eliminate confusion from overlapping families in our assessment of the characteristic properties of the Floras.
 \item We make use of reflectance information alongside orbital information, limiting the number of objects in our dataset to those for which both orbital and reflectance data exist. This step is necessary for the dynamically crowded Flora region, and not so injurious to the Floras due to the size of the family. 
 \item We use the signature of the family in $a$-H space as an additional tool to identify family members. This technique is not used to its full potential in the literature to date, and has led some to mistakenly separate single-collision families into two dynamical groups composed of the prograde and retrograde ``wings'' of the V plots.
 \item We establish the full range of dynamical space occupied by the Flora family, which can then be interpreted physically to obtain the age of the family, via observations of the dispersion in semimajor axis due to Yarkovsky spreading.
 \item We obtain a calibration for the Yarkovsky drift among the Floras via comparison with another S-type family, Karin. This assumes the close correspondence of Flora and Karin material properties, but enables the derivation of the Flora collisional age without physical assumptions about the members' material properties themselves, which are poorly constrained.
\end{itemize}

\subsection{Future Work}

Our definition of the ranges of the Flora family in orbital and reflectance properties provides an excellent opportunity to test the various models of asteroid spin evolution predicted by the models of YORP and spin-orbit resonance capture. Future efforts will model the $C$ distributions produced under various YORP and spin-orbit resonance capture scenarios, to understand the physical implications of the distribution on the family's evolution.

The Floras' properties are distinct from those of other overlapping or embedded asteroid families as shown in Table \ref{neighborhood}, based on preliminary evaluations of the properties of those other families.  One aspect of our future work will refine the results in this table, obtaining reliable identification of members of the various families packed into this region. With that information, the ages of the families and their collisional and dynamical histories can eventually be derived.

\begin{table}[H]
{\footnotesize
\begin{tabular}{ l | S c c c c S S S }
 \hline 
	Family name & {H} & $a$ & $e$ & $\sin i$ & $C$ envelope & {$a^*$} & {$i-z$} & {$p_{\text{V}}$} \\ [1.0ex]
 \hline
	27 Euterpe & 7.0 & 2.347 & 0.190 & 0.015 & 0.10 & 0.12 & -0.06 & 0.27 \\
	20 Massalia & 6.5 & 2.410 & 0.163 & 0.025 & 0.30 & 0.07 & -0.045 & 0.24 \\
	317 Roxane & 10.0 & 2.287 & 0.036 & 0.031 & 0.06 & -0.035 & 0.00 &  \\
	495 Eulalia & 10.8 & 2.487 & 0.110 & 0.040 & 0.12 & -0.13 & -0.01 & 0.057 \\
	495 Eulalia & 10.8 & 2.487 & 0.147 & 0.047 & 0.10 & -0.12 & -0.018 & 0.056 \\
	2751 Campbell & 12.8 & 2.410 & 0.186 & 0.040 & 0.05 & 0.12 & -0.03 & 0.29 \\
	135 Hertha & 8.2 & 2.427 & 0.170 & 0.044 & 0.05 & 0.13 & -0.035 & 0.28 \\
	135 Hertha & 8.2 & 2.427 & 0.175 & 0.045 & 0.01 & -0.075 & 0.045 & 0.18 \\
	302 Clarissa & 10.9 & 2.410 & 0.106 & 0.058 & 0.01 & -0.13 & -0.03 & 0.047 \\
	142 Polana & 10.3 & 2.418 & 0.155 & 0.059 & 0.32 & -0.12 & 0.02 & 0.058 \\
	554 Peraga & 9.0 & 2.375 & 0.175 & 0.060 & 0.08 &  &  &  \\
	8 Flora & 6.5 & 2.201 & 0.130 & 0.080 & 0.164 & 0.126 & -0.037 & 0.291 \\
	79 Eurynome & 8.0 & 2.444 & 0.174 & 0.084 & 0.04 & 0.16 & -0.04 & 0.28 \\
	163 Erigone & 12.5 & 2.369 & 0.206 & 0.087 & 0.02 & -0.09 & 0.05 & 0.05 \\
	752 Sulamitis & 10.2 & 2.463 & 0.090 & 0.088 & 0.03 & -0.07 & 0.07 & 0.051 \\
	298 Baptistina & 11.2 & 2.264 & 0.143 & 0.097 & 0.03 & -0.026 & 0.017 & 0.15 \\
	63 Ausonia & 7.6 & 2.395 & 0.110 & 0.110 & 0.14 & 0.11 & -0.25 & 0.30 \\
	67 Asia & 8.3 & 2.421 & 0.150 & 0.114 & 0.03 & 0.085 & -0.13 & 0.22 \\
	4 Vesta & 3.2 & 2.362 & 0.094 & 0.115 & 0.09 & 0.12 & -0.25 & 0.36 \\
	757 Portlandia & 10.2 & 2.373 & 0.105 & 0.136 & 0.20 & -0.035 & 0.04 & 0.18 \\
\hline
\end{tabular}

}
 \caption{Orbital and reflectance properties of neighboring families that overlap the Flora family (those with $\sin i < 0.15$, ordered here by increasing $\sin i$), results of a preliminary analysis using the method outlined in Section \ref{reflectance} above. The H and $a$ values are for the parent objects. The $C$ envelope, $a^*$ and $i-z$ colors, and $p_\text{V}$ albedo are defined in the text (equations \ref{Cparam}, \ref{astar_eq} and \ref{pV_eq}). Reflectance properties are not reported for those families whose membership is too small to obtain reliable estimates.}
 \label{neighborhood}
\end{table}

\section* {ACKNOWLEDGEMENTS}

{\scriptsize MJD was supported in part by an NSF Graduate Research Fellowship, Award No. DGE-1143953. SJV was supported by a Michigan Space Grant Consortium undergraduate fellowship and the Calvin College Integrated Science Research Institute. 

The authors acknowledge the use of data from the Sloan Digital Sky Survey, and thank the Sloan team and its sponsors (see http://www.sdss.org). In addition, this publication makes use of data products from the Wide-field Infrared Survey Explorer and NEOWISE, which is a joint project of the University of California, Los Angeles, and the Jet Propulsion Laboratory/California Institute of Technology, funded by the National Aeronautics and Space Administration.


\bibliographystyle{plainnat}
\bibliography{bibliography}

\begin{thebibliography}{60}
\providecommand{\natexlab}[1]{#1}
\providecommand{\url}[1]{\texttt{#1}}
\expandafter\ifx\csname urlstyle\endcsname\relax
  \providecommand{\doi}[1]{doi: #1}\else
  \providecommand{\doi}{doi: \begingroup \urlstyle{rm}\Url}\fi

\bibitem[{Bendjoya} and {Zappal{\`a}}(2002)]{bendjoya2002}
P.~{Bendjoya} and V.~{Zappal{\`a}}.
\newblock {Asteroid Family Identification}.
\newblock \emph{Asteroids III}, pages 613--618, 2002.

\bibitem[{Bendjoya} et~al.(1993){Bendjoya}, {Cellino}, {Froeschle}, and
  {Zappala}]{bendjoya1993}
P.~{Bendjoya}, A.~{Cellino}, C.~{Froeschle}, and V.~{Zappala}.
\newblock {Asteroid Dynamical Families: a Reliability Test for Two
  Identification Methods}.
\newblock \emph{\aap}, 272:\penalty0 651--670, May 1993.

\bibitem[{Bottke} et~al.(2013){Bottke}, {Vokrouhlicky}, {Nesvorny}, {Walsh},
  {Delbo}, {Lauretta}, {Connolly}, and {OSIRIS-REx Team}]{bottke2013}
W.~{Bottke}, D.~{Vokrouhlicky}, D.~{Nesvorny}, K.~{Walsh}, M.~{Delbo},
  D.~{Lauretta}, H.~{Connolly}, and {OSIRIS-REx Team}.
\newblock {The Unusual Evolution of Billion-Year Old Asteroid Families by the
  Yarkovsky and YORP Effects}.
\newblock In \emph{AAS/Division for Planetary Sciences Meeting Abstracts},
  volume~45 of \emph{AAS/Division for Planetary Sciences Meeting Abstracts},
  page \#106.06, October 2013.

\bibitem[{Bottke} et~al.(2001){Bottke}, {Vokrouhlick{\'y}}, {Broz},
  {Nesvorn{\'y}}, and {Morbidelli}]{bottke2001}
W.~F. {Bottke}, D.~{Vokrouhlick{\'y}}, M.~{Broz}, D.~{Nesvorn{\'y}}, and
  A.~{Morbidelli}.
\newblock {Dynamical Spreading of Asteroid Families by the Yarkovsky Effect}.
\newblock \emph{Science}, 294:\penalty0 1693--1696, November 2001.
\newblock \doi{10.1126/science.1066760}.

\bibitem[{Bottke} et~al.(2006){Bottke}, {Vokrouhlick{\'y}}, {Rubincam}, and
  {Nesvorn{\'y}}]{bottke2006}
W.~F. {Bottke}, Jr., D.~{Vokrouhlick{\'y}}, D.~P. {Rubincam}, and
  D.~{Nesvorn{\'y}}.
\newblock {The Yarkovsky and YORP Effects: Implications for Asteroid Dynamics}.
\newblock \emph{Annual Review of Earth and Planetary Sciences}, 34:\penalty0
  157--191, May 2006.
\newblock \doi{10.1146/annurev.earth.34.031405.125154}.

\bibitem[{Brouwer}(1951)]{brouwer1951}
D.~{Brouwer}.
\newblock {Secular variations of the orbital elements of minor planets}.
\newblock \emph{\aj}, 56:\penalty0 9, March 1951.
\newblock \doi{10.1086/106480}.

\bibitem[{Bro{\v z}} et~al.(2013){Bro{\v z}}, {Morbidelli}, {Bottke},
  {Rozehnal}, {Vokrouhlick{\'y}}, and {Nesvorn{\'y}}]{broz2013}
M.~{Bro{\v z}}, A.~{Morbidelli}, W.~F. {Bottke}, J.~{Rozehnal},
  D.~{Vokrouhlick{\'y}}, and D.~{Nesvorn{\'y}}.
\newblock {Constraining the cometary flux through the asteroid belt during the
  late heavy bombardment}.
\newblock \emph{\aap}, 551:\penalty0 A117, March 2013.
\newblock \doi{10.1051/0004-6361/201219296}.

\bibitem[{Bus} and {Binzel}(2002)]{bus2002}
S.~J. {Bus} and R.~P. {Binzel}.
\newblock {Phase II of the Small Main-Belt Asteroid Spectroscopic Survey. The
  Observations}.
\newblock \emph{Icarus}, 158:\penalty0 106--145, July 2002.
\newblock \doi{10.1006/icar.2002.6857}.

\bibitem[{Carruba} et~al.(2013){Carruba}, {Domingos}, {Nesvorn{\'y}}, {Roig},
  {Huaman}, and {Souami}]{carruba2013}
V.~{Carruba}, R.~C. {Domingos}, D.~{Nesvorn{\'y}}, F.~{Roig}, M.~E. {Huaman},
  and D.~{Souami}.
\newblock {A multidomain approach to asteroid families' identification}.
\newblock \emph{\mnras}, 433:\penalty0 2075--2096, August 2013.
\newblock \doi{10.1093/mnras/stt884}.

\bibitem[{Carry}(2012)]{carry2012}
B.~{Carry}.
\newblock {Density of asteroids}.
\newblock \emph{\planss}, 73:\penalty0 98--118, December 2012.
\newblock \doi{10.1016/j.pss.2012.03.009}.

\bibitem[{Carusi} and {Massaro}(1978)]{carusi1978}
A.~{Carusi} and E.~{Massaro}.
\newblock {Statistics and mapping of asteroid concentrations in the proper
  elements space}.
\newblock \emph{\aaps}, 34:\penalty0 81--90, October 1978.

\bibitem[{Cellino} et~al.(1991){Cellino}, {Zappala}, and
  {Farinella}]{cellino1991}
A.~{Cellino}, V.~{Zappala}, and P.~{Farinella}.
\newblock {The size distribution of main-belt asteroids from IRAS data}.
\newblock \emph{\mnras}, 253:\penalty0 561--574, December 1991.

\bibitem[{Cellino} et~al.(2009){Cellino}, {Dell'Oro}, and
  {Tedesco}]{cellino2009}
A.~{Cellino}, A.~{Dell'Oro}, and E.~F. {Tedesco}.
\newblock {Asteroid families: Current situation}.
\newblock \emph{\planss}, 57:\penalty0 173--182, February 2009.
\newblock \doi{10.1016/j.pss.2008.07.028}.

\bibitem[{Chapman}(1996)]{chapman1996}
C.~R. {Chapman}.
\newblock {S-Type Asteroids, Ordinary Chondrites, and Space Weathering: The
  Evidence from Galileo's Fly-bys of Gaspra and Ida}.
\newblock \emph{Meteoritics and Planetary Science}, 31:\penalty0 699--725,
  November 1996.
\newblock \doi{10.1111/j.1945-5100.1996.tb02107.x}.

\bibitem[Conover(1980)]{conover1980}
W.~J. Conover.
\newblock Practical nonparametric statistics.
\newblock 1980.

\bibitem[{Cotto-Figueroa} et~al.(2013){Cotto-Figueroa}, {Statler},
  {Richardson}, and {Tanga}]{cotto-figueroa2013}
D.~{Cotto-Figueroa}, T.~S. {Statler}, D.~C. {Richardson}, and P.~{Tanga}.
\newblock {Killing the YORP Cycle: A Stochastic and Self-Limiting YORP Effect}.
\newblock In \emph{AAS/Division for Planetary Sciences Meeting Abstracts},
  volume~45 of \emph{AAS/Division for Planetary Sciences Meeting Abstracts},
  page \#106.09, October 2013.

\bibitem[Fowler and Chillemi(1992)]{fowler1992}
J.W. Fowler and J.R. Chillemi.
\newblock Iras asteroid data processing.
\newblock \emph{The IRAS Minor Planet Survey. Phillips Laboratory, Hanscom Air
  Force Base, MA}, pages 17--43, 1992.

\bibitem[{Fukugita} et~al.(1996){Fukugita}, {Ichikawa}, {Gunn}, {Doi},
  {Shimasaku}, and {Schneider}]{fukugita1996}
M.~{Fukugita}, T.~{Ichikawa}, J.~E. {Gunn}, M.~{Doi}, K.~{Shimasaku}, and D.~P.
  {Schneider}.
\newblock {The Sloan Digital Sky Survey Photometric System}.
\newblock \emph{\aj}, 111:\penalty0 1748, April 1996.
\newblock \doi{10.1086/117915}.

\bibitem[{Greenberg} et~al.(1994){Greenberg}, {Nolan}, {Bottke}, {Kolvoord},
  and {Veverka}]{greenberg1994}
R.~{Greenberg}, M.~C. {Nolan}, W.~F. {Bottke}, Jr., R.~A. {Kolvoord}, and
  J.~{Veverka}.
\newblock {Collisional history of Gaspra}.
\newblock \emph{Icarus}, 107:\penalty0 84, January 1994.
\newblock \doi{10.1006/icar.1994.1008}.

\bibitem[{Haegert} and {Molnar}(2009)]{haegert2009}
M.~J. {Haegert} and L.~A. {Molnar}.
\newblock {Flora Spin Axis Survey: Confirmation of the Yarkovsky Effect}.
\newblock In \emph{AAS/Division for Planetary Sciences Meeting Abstracts \#41},
  volume~41 of \emph{AAS/Division for Planetary Sciences Meeting Abstracts},
  page \#56.02, September 2009.

\bibitem[{Hanu{\v s}} et~al.(2013){Hanu{\v s}}, {Bro{\v z}}, {{\v D}urech},
  {Warner}, {Brinsfield}, {Durkee}, {Higgins}, {Koff}, {Oey}, {Pilcher},
  {Stephens}, {Strabla}, {Ulisse}, and {Girelli}]{hanus2013}
J.~{Hanu{\v s}}, M.~{Bro{\v z}}, J.~{{\v D}urech}, B.~D. {Warner},
  J.~{Brinsfield}, R.~{Durkee}, D.~{Higgins}, R.~A. {Koff}, J.~{Oey},
  F.~{Pilcher}, R.~{Stephens}, L.~P. {Strabla}, Q.~{Ulisse}, and R.~{Girelli}.
\newblock {An anisotropic distribution of spin vectors in asteroid families}.
\newblock \emph{ArXiv e-prints}, September 2013.

\bibitem[{Hirayama}(1919)]{hirayama1919}
K.~{Hirayama}.
\newblock {Further note on the families of asteroids}.
\newblock \emph{Annales de l'Observatoire astronomique de Tokyo}, 8:\penalty0
  A1, 1919.

\bibitem[{Ivezi{\'c}} et~al.(2001){Ivezi{\'c}}, {Tabachnik}, {Rafikov},
  {Lupton}, {Quinn}, {Hammergren}, {Eyer}, {Chu}, {Armstrong}, {Fan},
  {Finlator}, {Geballe}, {Gunn}, {Hennessy}, {Knapp}, {Leggett}, {Munn},
  {Pier}, {Rockosi}, {Schneider}, {Strauss}, {Yanny}, {Brinkmann}, {Csabai},
  {Hindsley}, {Kent}, {Lamb}, {Margon}, {McKay}, {Smith}, {Waddel}, {York}, and
  {SDSS Collaboration}]{ivezic2001}
{\v Z}.~{Ivezi{\'c}}, S.~{Tabachnik}, R.~{Rafikov}, R.~H. {Lupton}, T.~{Quinn},
  M.~{Hammergren}, L.~{Eyer}, J.~{Chu}, J.~C. {Armstrong}, X.~{Fan},
  K.~{Finlator}, T.~R. {Geballe}, J.~E. {Gunn}, G.~S. {Hennessy}, G.~R.
  {Knapp}, S.~K. {Leggett}, J.~A. {Munn}, J.~R. {Pier}, C.~M. {Rockosi}, D.~P.
  {Schneider}, M.~A. {Strauss}, B.~{Yanny}, J.~{Brinkmann}, I.~{Csabai}, R.~B.
  {Hindsley}, S.~{Kent}, D.~Q. {Lamb}, B.~{Margon}, T.~A. {McKay}, J.~A.
  {Smith}, P.~{Waddel}, D.~G. {York}, and {SDSS Collaboration}.
\newblock {Solar System Objects Observed in the Sloan Digital Sky Survey
  Commissioning Data}.
\newblock \emph{\aj}, 122:\penalty0 2749--2784, November 2001.
\newblock \doi{10.1086/323452}.

\bibitem[{Ivezic} et~al.(2002){Ivezic}, {Juric}, {Lupton}, {Tabachnik}, and
  {Quinn}]{ivezic2002}
Z.~{Ivezic}, M.~{Juric}, R.~H. {Lupton}, S.~{Tabachnik}, and T.~{Quinn}.
\newblock {Asteroids Observed by The Sloan Digital Survey}.
\newblock In J.~A. {Tyson} and S.~{Wolff}, editors, \emph{Survey and Other
  Telescope Technologies and Discoveries}, volume 4836 of \emph{Society of
  Photo-Optical Instrumentation Engineers (SPIE) Conference Series}, pages
  98--103, December 2002.
\newblock \doi{10.1117/12.457304}.

\bibitem[{Konopliv} et~al.(2011){Konopliv}, {Asmar}, {Folkner}, {Karatekin},
  {Nunes}, {Smrekar}, {Yoder}, and {Zuber}]{konopliv2011}
A.~S. {Konopliv}, S.~W. {Asmar}, W.~M. {Folkner}, {\"O}.~{Karatekin}, D.~C.
  {Nunes}, S.~E. {Smrekar}, C.~F. {Yoder}, and M.~T. {Zuber}.
\newblock {Mars high resolution gravity fields from MRO, Mars seasonal gravity,
  and other dynamical parameters}.
\newblock \emph{Icarus}, 211:\penalty0 401--428, January 2011.
\newblock \doi{10.1016/j.icarus.2010.10.004}.

\bibitem[{Kryszczy{\'n}ska}(2013)]{kryszczynska2013}
A.~{Kryszczy{\'n}ska}.
\newblock {Do Slivan states exist in the Flora family? . II. Fingerprints of
  the Yarkovsky and YORP effects}.
\newblock \emph{\aap}, 551:\penalty0 A102, March 2013.
\newblock \doi{10.1051/0004-6361/201220490}.

\bibitem[{Mainzer} et~al.(2011){Mainzer}, {Bauer}, {Grav}, {Masiero}, {Cutri},
  {Dailey}, {Eisenhardt}, {McMillan}, {Wright}, {Walker}, {Jedicke}, {Spahr},
  {Tholen}, {Alles}, {Beck}, {Brandenburg}, {Conrow}, {Evans}, {Fowler},
  {Jarrett}, {Marsh}, {Masci}, {McCallon}, {Wheelock}, {Wittman}, {Wyatt},
  {DeBaun}, {Elliott}, {Elsbury}, {Gautier}, {Gomillion}, {Leisawitz},
  {Maleszewski}, {Micheli}, and {Wilkins}]{mainzer2011}
A.~{Mainzer}, J.~{Bauer}, T.~{Grav}, J.~{Masiero}, R.~M. {Cutri}, J.~{Dailey},
  P.~{Eisenhardt}, R.~S. {McMillan}, E.~{Wright}, R.~{Walker}, R.~{Jedicke},
  T.~{Spahr}, D.~{Tholen}, R.~{Alles}, R.~{Beck}, H.~{Brandenburg},
  T.~{Conrow}, T.~{Evans}, J.~{Fowler}, T.~{Jarrett}, K.~{Marsh}, F.~{Masci},
  H.~{McCallon}, S.~{Wheelock}, M.~{Wittman}, P.~{Wyatt}, E.~{DeBaun},
  G.~{Elliott}, D.~{Elsbury}, T.~{Gautier}, IV, S.~{Gomillion}, D.~{Leisawitz},
  C.~{Maleszewski}, M.~{Micheli}, and A.~{Wilkins}.
\newblock {Preliminary Results from NEOWISE: An Enhancement to the Wide-field
  Infrared Survey Explorer for Solar System Science}.
\newblock \emph{\apj}, 731:\penalty0 53, April 2011.
\newblock \doi{10.1088/0004-637X/731/1/53}.

\bibitem[{Marchi} et~al.(){Marchi}, {Bottke}, {O'Brien}, {Schenk}, {Mottola},
  {De Sanctis}, {Kring}, {Williams}, {Raymond}, and {Russell}]{marchi2013}
S.~{Marchi}, W.~F. {Bottke}, D.~P. {O'Brien}, P.~{Schenk}, S.~{Mottola}, M.~C.
  {De Sanctis}, D.~A. {Kring}, D.~A. {Williams}, C.~A. {Raymond}, and C.~T.
  {Russell}.
\newblock {Small crater populations on Vesta}.
\newblock \emph{\planss ~(in press)}.

\bibitem[{Masiero} et~al.(2011){Masiero}, {Mainzer}, {Grav}, {Bauer}, {Cutri},
  {Dailey}, {Eisenhardt}, {McMillan}, {Spahr}, {Skrutskie}, {Tholen}, {Walker},
  {Wright}, {DeBaun}, {Elsbury}, {Gautier}, {Gomillion}, and
  {Wilkins}]{masiero2011}
J.~R. {Masiero}, A.~K. {Mainzer}, T.~{Grav}, J.~M. {Bauer}, R.~M. {Cutri},
  J.~{Dailey}, P.~R.~M. {Eisenhardt}, R.~S. {McMillan}, T.~B. {Spahr}, M.~F.
  {Skrutskie}, D.~{Tholen}, R.~G. {Walker}, E.~L. {Wright}, E.~{DeBaun},
  D.~{Elsbury}, T.~{Gautier}, IV, S.~{Gomillion}, and A.~{Wilkins}.
\newblock {Main Belt Asteroids with WISE/NEOWISE. I. Preliminary Albedos and
  Diameters}.
\newblock \emph{\apj}, 741:\penalty0 68, November 2011.
\newblock \doi{10.1088/0004-637X/741/2/68}.

\bibitem[{Masiero} et~al.(2013){Masiero}, {Mainzer}, {Bauer}, {Grav}, {Nugent},
  and {Stevenson}]{masiero2013}
J.~R. {Masiero}, A.~K. {Mainzer}, J.~M. {Bauer}, T.~{Grav}, C.~R. {Nugent}, and
  R.~{Stevenson}.
\newblock {Asteroid Family Identification Using the Hierarchical Clustering
  Method and WISE/NEOWISE Physical Properties}.
\newblock \emph{\apj}, 770:\penalty0 7, June 2013.
\newblock \doi{10.1088/0004-637X/770/1/7}.

\bibitem[{Milani} et~al.(2013){Milani}, {Cellino}, {Knezevic}, {Novakovic},
  {Spoto}, and {Paolicchi}]{milani2013}
A.~{Milani}, A.~{Cellino}, Z.~{Knezevic}, B.~{Novakovic}, F.~{Spoto}, and
  P.~{Paolicchi}.
\newblock {Asteroid families classification: exploiting very large data sets}.
\newblock \emph{ArXiv e-prints}, December 2013.

\bibitem[{Molnar}(2011)]{molnar2011}
L.~A. {Molnar}.
\newblock {Size and Age Dependence of Koronis Family Colors}.
\newblock In \emph{EPSC-DPS Joint Meeting 2011}, page 1684, October 2011.

\bibitem[{Molnar} and {Haegert}(2008)]{molnar2008}
L.~A. {Molnar} and M.~J. {Haegert}.
\newblock {The Yarkovsky Effect in the Flora and Baptistina Asteroid Families}.
\newblock In \emph{AAS/Division of Dynamical Astronomy Meeting \#39}, volume~39
  of \emph{AAS/Division of Dynamical Astronomy Meeting}, page \#02.03, May
  2008.

\bibitem[{Morbidelli} and {Vokrouhlick{\'y}}(2003)]{morbidelli2003}
A.~{Morbidelli} and D.~{Vokrouhlick{\'y}}.
\newblock {The Yarkovsky-driven origin of near-Earth asteroids}.
\newblock \emph{Icarus}, 163:\penalty0 120--134, May 2003.
\newblock \doi{10.1016/S0019-1035(03)00047-2}.

\bibitem[{Nesvorn{\'y}}(2012)]{nesvorny2012}
D.~{Nesvorn{\'y}}.
\newblock {Nesvorn{\'y} HCM Asteroid Families V2.0}.
\newblock \emph{NASA Planetary Data System}, 189, June 2012.

\bibitem[{Nesvorn{\'y}} and {Bottke}(2004)]{nesvorny2004}
D.~{Nesvorn{\'y}} and W.~F. {Bottke}.
\newblock {Detection of the Yarkovsky effect for main-belt asteroids}.
\newblock \emph{Icarus}, 170:\penalty0 324--342, August 2004.
\newblock \doi{10.1016/j.icarus.2004.04.012}.

\bibitem[{Nesvorn{\'y}} and {Vokrouhlick{\'y}}(2008)]{nesvorny2008}
D.~{Nesvorn{\'y}} and D.~{Vokrouhlick{\'y}}.
\newblock {Vanishing torque from radiation pressure}.
\newblock \emph{\aap}, 480:\penalty0 1--3, March 2008.
\newblock \doi{10.1051/0004-6361:20078389}.

\bibitem[{Nesvorn{\'y}} et~al.(2002){Nesvorn{\'y}}, {Morbidelli},
  {Vokrouhlick{\'y}}, {Bottke}, and {Bro{\v z}}]{nesvorny2002}
D.~{Nesvorn{\'y}}, A.~{Morbidelli}, D.~{Vokrouhlick{\'y}}, W.~F. {Bottke}, and
  M.~{Bro{\v z}}.
\newblock {The Flora Family: A Case of the Dynamically Dispersed Collisional
  Swarm?}
\newblock \emph{Icarus}, 157:\penalty0 155--172, May 2002.
\newblock \doi{10.1006/icar.2002.6830}.

\bibitem[{Nesvorn{\'y}} et~al.(2005){Nesvorn{\'y}}, {Jedicke}, {Whiteley}, and
  {Ivezi{\'c}}]{nesvorny2005}
D.~{Nesvorn{\'y}}, R.~{Jedicke}, R.~J. {Whiteley}, and {\v Z}.~{Ivezi{\'c}}.
\newblock {Evidence for asteroid space weathering from the Sloan Digital Sky
  Survey}.
\newblock \emph{Icarus}, 173:\penalty0 132--152, January 2005.
\newblock \doi{10.1016/j.icarus.2004.07.026}.

\bibitem[{Nesvorn{\'y}} et~al.(2007){Nesvorn{\'y}}, {Vokrouhlick{\'y}},
  {Bottke}, {Gladman}, and {H{\"a}ggstr{\"o}m}]{nesvorny2007}
D.~{Nesvorn{\'y}}, D.~{Vokrouhlick{\'y}}, W.~F. {Bottke}, B.~{Gladman}, and
  T.~{H{\"a}ggstr{\"o}m}.
\newblock {Express delivery of fossil meteorites from the inner asteroid belt
  to Sweden}.
\newblock \emph{Icarus}, 188:\penalty0 400--413, June 2007.
\newblock \doi{10.1016/j.icarus.2006.11.021}.

\bibitem[{Nugent} et~al.(2012){Nugent}, {Mainzer}, {Masiero}, {Grav}, and
  {Bauer}]{nugent2012}
C.~R. {Nugent}, A.~{Mainzer}, J.~{Masiero}, T.~{Grav}, and J.~{Bauer}.
\newblock {The Yarkovsky Drift's Influence on NEAs: Trends and Predictions with
  NEOWISE Measurements}.
\newblock \emph{\aj}, 144:\penalty0 75, September 2012.
\newblock \doi{10.1088/0004-6256/144/3/75}.

\bibitem[{O'Brien} et~al.(2006){O'Brien}, {Greenberg}, and
  {Richardson}]{obrien2006}
D.~P. {O'Brien}, R.~{Greenberg}, and J.~E. {Richardson}.
\newblock {Craters on asteroids: Reconciling diverse impact records with a
  common impacting population}.
\newblock \emph{Icarus}, 183:\penalty0 79--92, July 2006.
\newblock \doi{10.1016/j.icarus.2006.02.008}.

\bibitem[{Parker} et~al.(2008){Parker}, {Ivezi{\'c}}, {Juri{\'c}}, {Lupton},
  {Sekora}, and {Kowalski}]{parker2008}
A.~{Parker}, {\v Z}.~{Ivezi{\'c}}, M.~{Juri{\'c}}, R.~{Lupton}, M.~D. {Sekora},
  and A.~{Kowalski}.
\newblock {The size distributions of asteroid families in the SDSS Moving
  Object Catalog 4}.
\newblock \emph{Icarus}, 198:\penalty0 138--155, November 2008.
\newblock \doi{10.1016/j.icarus.2008.07.002}.

\bibitem[{Reddy} et~al.(2011){Reddy}, {Carvano}, {Lazzaro}, {Michtchenko},
  {Gaffey}, {Kelley}, {Moth{\'e}-Diniz}, {Alvarez-Candal}, {Moskovitz},
  {Cloutis}, and {Ryan}]{reddy2011}
V.~{Reddy}, J.~M. {Carvano}, D.~{Lazzaro}, T.~A. {Michtchenko}, M.~J. {Gaffey},
  M.~S. {Kelley}, T.~{Moth{\'e}-Diniz}, A.~{Alvarez-Candal}, N.~A. {Moskovitz},
  E.~A. {Cloutis}, and E.~L. {Ryan}.
\newblock {Mineralogical characterization of Baptistina Asteroid Family:
  Implications for K/T impactor source}.
\newblock \emph{Icarus}, 216:\penalty0 184--197, November 2011.
\newblock \doi{10.1016/j.icarus.2011.08.027}.

\bibitem[{Rubincam} et~al.(2002){Rubincam}, {Rowlands}, and
  {Ray}]{rubincam2002}
D.~P. {Rubincam}, D.~D. {Rowlands}, and R.~D. {Ray}.
\newblock {Is asteroid 951 Gaspra in a resonant obliquity state with its spin
  increasing due to YORP?}
\newblock \emph{Journal of Geophysical Research (Planets)}, 107:\penalty0 5065,
  September 2002.
\newblock \doi{10.1029/2001JE001813}.

\bibitem[{Scholl} and {Froeschle}(1991)]{scholl1991}
H.~{Scholl} and C.~{Froeschle}.
\newblock {The nu(6) secular resonance region near 2 AU - A possible source of
  meteorites}.
\newblock \emph{\aap}, 245:\penalty0 316--321, May 1991.

\bibitem[{Slivan} et~al.(2003){Slivan}, {Binzel}, {Crespo da Silva},
  {Kaasalainen}, {Lyndaker}, and {Kr{\v c}o}]{slivan2003}
S.~M. {Slivan}, R.~P. {Binzel}, L.~D. {Crespo da Silva}, M.~{Kaasalainen},
  M.~M. {Lyndaker}, and M.~{Kr{\v c}o}.
\newblock {Spin vectors in the Koronis family: comprehensive results from two
  independent analyses of 213 rotation lightcurves}.
\newblock \emph{Icarus}, 162:\penalty0 285--307, April 2003.
\newblock \doi{10.1016/S0019-1035(03)00029-0}.

\bibitem[{Statler}(2009)]{statler2009}
T.~S. {Statler}.
\newblock {Extreme sensitivity of the YORP effect to small-scale topography}.
\newblock \emph{Icarus}, 202:\penalty0 502--513, August 2009.
\newblock \doi{10.1016/j.icarus.2009.03.003}.

\bibitem[{Tedesco}(1979)]{tedesco1979}
E.~F. {Tedesco}.
\newblock {The origin of the Flora family}.
\newblock \emph{Icarus}, 40:\penalty0 375--382, December 1979.
\newblock \doi{10.1016/0019-1035(79)90030-7}.

\bibitem[{Vernazza} et~al.(2008){Vernazza}, {Binzel}, {Thomas}, {DeMeo}, {Bus},
  {Rivkin}, and {Tokunaga}]{vernazza2008}
P.~{Vernazza}, R.~P. {Binzel}, C.~A. {Thomas}, F.~E. {DeMeo}, S.~J. {Bus},
  A.~S. {Rivkin}, and A.~T. {Tokunaga}.
\newblock {Compositional differences between meteorites and near-Earth
  asteroids}.
\newblock \emph{\nat}, 454:\penalty0 858--860, August 2008.
\newblock \doi{10.1038/nature07154}.

\bibitem[{Vokrouhlick{\'y}} and {Farinella}(2000)]{vokrouhlicky2000}
D.~{Vokrouhlick{\'y}} and P.~{Farinella}.
\newblock {Efficient delivery of meteorites to the Earth from a wide range of
  asteroid parent bodies}.
\newblock \emph{\nat}, 407:\penalty0 606--608, October 2000.

\bibitem[{Vokrouhlick{\'y}} and {{\v C}apek}(2002)]{vokrouhlicky2002}
D.~{Vokrouhlick{\'y}} and D.~{{\v C}apek}.
\newblock {YORP-Induced Long-Term Evolution of the Spin State of Small
  Asteroids and Meteoroids: Rubincam's Approximation}.
\newblock \emph{Icarus}, 159:\penalty0 449--467, October 2002.
\newblock \doi{10.1006/icar.2002.6918}.

\bibitem[{Vokrouhlick{\'y}} et~al.(2003){Vokrouhlick{\'y}}, {Nesvorn{\'y}}, and
  {Bottke}]{vokrouhlicky2003}
D.~{Vokrouhlick{\'y}}, D.~{Nesvorn{\'y}}, and W.~F. {Bottke}.
\newblock {The vector alignments of asteroid spins by thermal torques}.
\newblock \emph{\nat}, 425:\penalty0 147--151, September 2003.

\bibitem[{Vokrouhlick{\'y}} et~al.(2006{\natexlab{a}}){Vokrouhlick{\'y}},
  {Bro{\v z}}, {Bottke}, {Nesvorn{\'y}}, and {Morbidelli}]{vokrouhlicky2006}
D.~{Vokrouhlick{\'y}}, M.~{Bro{\v z}}, W.~F. {Bottke}, D.~{Nesvorn{\'y}}, and
  A.~{Morbidelli}.
\newblock {Yarkovsky/YORP chronology of asteroid families}.
\newblock \emph{Icarus}, 182:\penalty0 118--142, May 2006{\natexlab{a}}.
\newblock \doi{10.1016/j.icarus.2005.12.010}.

\bibitem[{Vokrouhlick{\'y}} et~al.(2006{\natexlab{b}}){Vokrouhlick{\'y}},
  {Nesvorn{\'y}}, and {Bottke}]{vokrouhlicky2006b}
D.~{Vokrouhlick{\'y}}, D.~{Nesvorn{\'y}}, and W.~F. {Bottke}.
\newblock {Secular spin dynamics of inner main-belt asteroids}.
\newblock \emph{Icarus}, 184:\penalty0 1--28, September 2006{\natexlab{b}}.
\newblock \doi{10.1016/j.icarus.2006.04.007}.

\bibitem[{Walsh} et~al.(2008){Walsh}, {Richardson}, and {Michel}]{walsh2008}
K.~J. {Walsh}, D.~C. {Richardson}, and P.~{Michel}.
\newblock {Rotational breakup as the origin of small binary asteroids}.
\newblock \emph{\nat}, 454:\penalty0 188--191, July 2008.
\newblock \doi{10.1038/nature07078}.

\bibitem[{Wright} et~al.(2010){Wright}, {Eisenhardt}, {Mainzer}, {Ressler},
  {Cutri}, {Jarrett}, {Kirkpatrick}, {Padgett}, {McMillan}, {Skrutskie},
  {Stanford}, {Cohen}, {Walker}, {Mather}, {Leisawitz}, {Gautier}, {McLean},
  {Benford}, {Lonsdale}, {Blain}, {Mendez}, {Irace}, {Duval}, {Liu}, {Royer},
  {Heinrichsen}, {Howard}, {Shannon}, {Kendall}, {Walsh}, {Larsen}, {Cardon},
  {Schick}, {Schwalm}, {Abid}, {Fabinsky}, {Naes}, and {Tsai}]{wright2010}
E.~L. {Wright}, P.~R.~M. {Eisenhardt}, A.~K. {Mainzer}, M.~E. {Ressler}, R.~M.
  {Cutri}, T.~{Jarrett}, J.~D. {Kirkpatrick}, D.~{Padgett}, R.~S. {McMillan},
  M.~{Skrutskie}, S.~A. {Stanford}, M.~{Cohen}, R.~G. {Walker}, J.~C. {Mather},
  D.~{Leisawitz}, T.~N. {Gautier}, III, I.~{McLean}, D.~{Benford}, C.~J.
  {Lonsdale}, A.~{Blain}, B.~{Mendez}, W.~R. {Irace}, V.~{Duval}, F.~{Liu},
  D.~{Royer}, I.~{Heinrichsen}, J.~{Howard}, M.~{Shannon}, M.~{Kendall}, A.~L.
  {Walsh}, M.~{Larsen}, J.~G. {Cardon}, S.~{Schick}, M.~{Schwalm}, M.~{Abid},
  B.~{Fabinsky}, L.~{Naes}, and C.-W. {Tsai}.
\newblock {The Wide-field Infrared Survey Explorer (WISE): Mission Description
  and Initial On-orbit Performance}.
\newblock \emph{\aj}, 140:\penalty0 1868, December 2010.
\newblock \doi{10.1088/0004-6256/140/6/1868}.

\bibitem[{Xu} et~al.(1995){Xu}, {Binzel}, {Burbine}, and {Bus}]{xu1995}
S.~{Xu}, R.~P. {Binzel}, T.~H. {Burbine}, and S.~J. {Bus}.
\newblock {Small main-belt asteroid spectroscopic survey: Initial results}.
\newblock \emph{Icarus}, 115:\penalty0 1--35, May 1995.
\newblock \doi{10.1006/icar.1995.1075}.

\bibitem[{Zappala} et~al.(1990){Zappala}, {Cellino}, {Farinella}, and
  {Knezevic}]{zappala1990}
V.~{Zappala}, A.~{Cellino}, P.~{Farinella}, and Z.~{Knezevic}.
\newblock {Asteroid families. I - Identification by hierarchical clustering and
  reliability assessment}.
\newblock \emph{\aj}, 100:\penalty0 2030--2046, December 1990.
\newblock \doi{10.1086/115658}.

\bibitem[{Zappala} et~al.(1994){Zappala}, {Cellino}, {Farinella}, and
  {Milani}]{zappala1994}
V.~{Zappala}, A.~{Cellino}, P.~{Farinella}, and A.~{Milani}.
\newblock {Asteroid families. 2: Extension to unnumbered multiopposition
  asteroids}.
\newblock \emph{\aj}, 107:\penalty0 772--801, February 1994.
\newblock \doi{10.1086/116897}.

\end{thebibliography}
}
\clearpage

\clearpage

\appendix

\section{Appendix A: Calculation of Confidence Interval for the Population Median}
\label{appendixA}

We use the following procedure from standard statistics texts \citep[e.g.,][]{conover1980} to calculate a confidence interval for the population median $M$ based on a sample of objects from that population. The following assumes that the sample selected is representative of the underlying population, and that the number of objects in the sample is large (N $>$ $\sim25$). Let the ordered data be denoted by $y_1,y_2,...,y_N$. A (1-$\alpha$) confidence interval for $M$ is given by the formula:

\begin{equation}
 (M_L,M_U) = (y_{C_\alpha+1},y_{N-C_\alpha}),
\end{equation}

\noindent where $C_\alpha$ is the appropriate quantile from the binomial distribution, but since N is always large in the cases where we apply this formula, a normal approximation is appropriate: 

\begin{equation}
 C_\alpha = \frac{N}{2} - z_\alpha \frac{\sqrt{N}}{2}.
\end{equation}

The value of $z_\alpha$ (1, 2, 3, ...) determines the confidence level (68\%, 95\%, 99\%) for the interval. For our estimate of 1$\sigma$ error bars, we set $z_\alpha = 1$, and expect that the true population median will be within the cited confidence interval 68\% of the time.

The assumption that the sample is representative of the population is only valid in cases where the method of selecting the sample does not bias the variable for which the median is determined. In general, this should be the case, as we are careful to make selections of objects only on the basis of their characteristics in all other parameters except the one under consideration. The nature of the original collision is likely such that some of the parameters are correlated (i.e., the family's distribution in proper element space exhibits the standard triaxial ellipsoid characteristic of collisional ejecta). This would imply that selections in one dimension (such as eccentricity) would affect the true underlying population's distribution in another dimension (such as inclination); however, we expect the selections to leave unaffected the symmetry of the distribution, thereby not biasing the median. The effect of the selections on the confidence interval in this case would be to tighten the interval slightly; thus we would expect the true interval to be slightly wider than the values obtained via the above method.

\section{Appendix B: Uniformity of Reflectance Properties near the Center in $e$ and $\sin i$}
\label{appendixB}

Here we explore the uniformity of the reflectance properties near the dynamical core of the Flora family, to assess the sensitivity of our characteristic reflectance properties to the uncertainty in the dynamical center in $e$ and $\sin i$ and use this sensitivity to define a reasonable range of uncertainty on the characteristic reflectance properties. 

We consider the uniformity of the reflectance properties within a region defined by the uncertainty on the center: $0.128 < e < 0.132$, $0.078 < \sin i < 0.082$. Within this region, we evaluate the median reflectance properties at each point within a 20 x 20 grid of ``pixels,'' or new center locations, mapped onto the region (400 pixels in all); that is to say, at each pixel we find the median $a^*$, $i-z$ and $p_\text{V}$ of all of the objects in a box centered on that pixel. The size of the box is a free parameter which we choose to be large enough to contain a minimum of about 30 objects (to avoid issues with the statistical analysis) and small enough to avoid significant contamination from nearby families. The box size that best met these criteria was $\pm$ 0.007 in both $e$ and $\sin i$ (i.e., a box of dimensions 0.014 x 0.014 in $e$ and $\sin i$); however, boxes with sizes within the range of $\pm$ 0.004 and $\pm$ 0.01 did not change the median reflectance properties beyond the uncertainties.

\begin{figure}
\begin{center}
\includegraphics [width=2.9in]{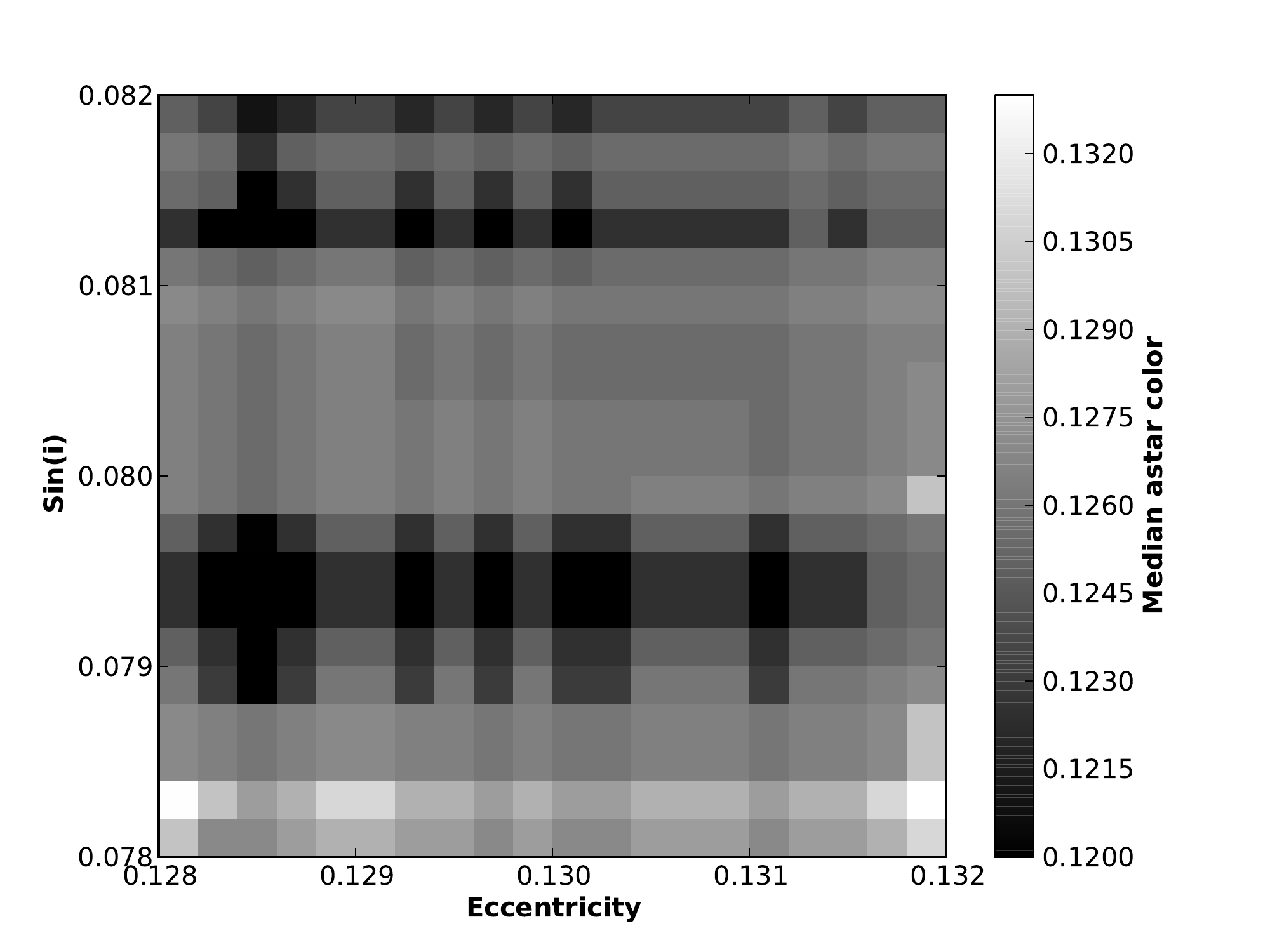}
\includegraphics [width=2.9in]{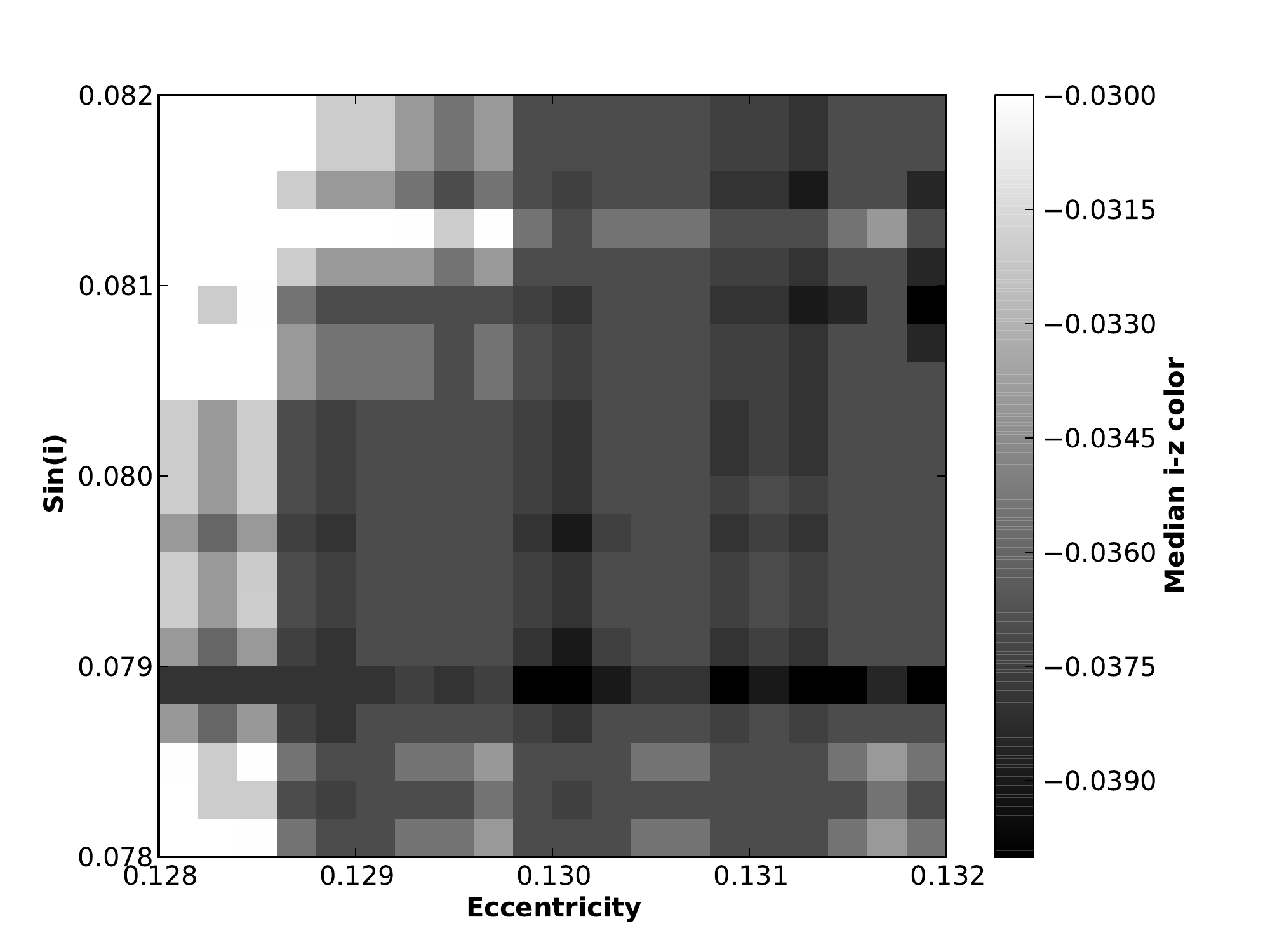}
\includegraphics [width=2.9in]{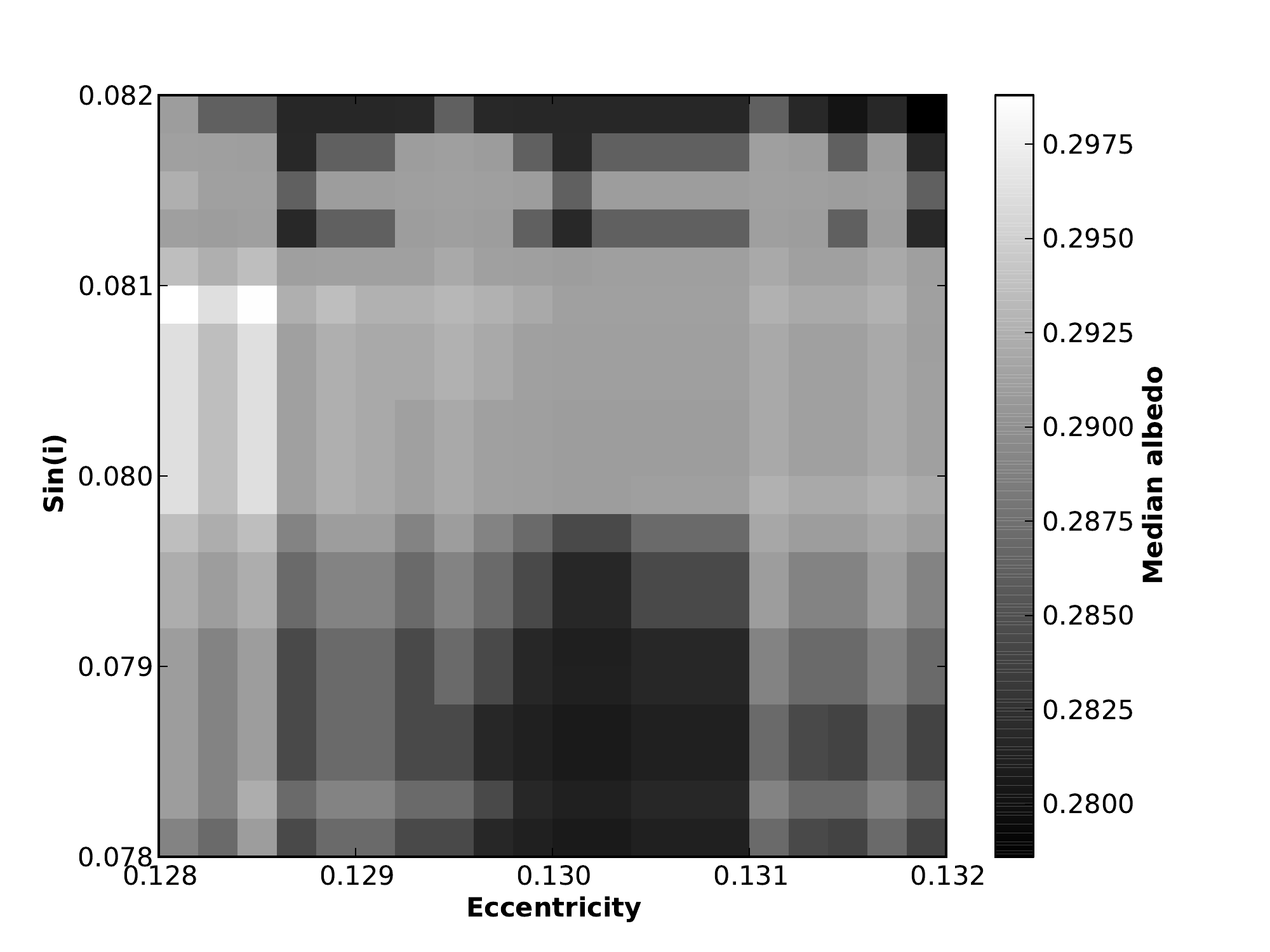}
\caption {\scriptsize Uniformity of the three reflectance parameters in the center region. To create the plot, we placed a 20x20 grid over the region defined by the uncertainty on the center in $e$ and $\sin i$, and for each ``pixel'' in this grid, we computed the median of all of the objects within a box with range $\pm$ 0.007 in both $e$ and $\sin i$ around that pixel. For each reflectance parameter, we report the medians of all of the pixels as the characteristic value for the family, and the range of all of the pixels as the uncertainty on that characteristic value. 
\label{A1_flatness}}
\end{center}
\end{figure}

The uniformity of the reflectance properties in the center region for a box size of $\pm$ 0.007 is plotted in Figure \ref{A1_flatness}. 

\end{document}